\documentclass[a4paper,fleqn,usenatbib,useAMS]{mnras}

\usepackage{graphicx}	
\usepackage{amsmath}	
\usepackage{amssymb}	
\usepackage{multicol}        
\usepackage{bm}		
\usepackage{pdflscape}	

\usepackage[T1]{fontenc}
\usepackage{ae,aecompl}

\usepackage{times,txfonts}

\title[Helioseismology of sunspot models]{$f$-mode interaction with models of sunspot : near-field scattering and multi-frequency effects}

\author[K. Daiffallah]{Khalil. Daiffallah$^{1}$\thanks{Contact e-mail: \href{mailto:k.daiffallah@craag.dz}{k.daiffallah@craag.dz}}
\\
$^{1}$Observatory of Algiers, CRAAG, Route de l'Observatoire, BP 63, Bouzar\'eah 16340, Algiers, Algeria}

\date{Last updated 2015 May 22; in original form 2013 September 5}

\pubyear{2016}

\begin{document}
\label{firstpage}
\pagerange{\pageref{firstpage}--\pageref{lastpage}}
\maketitle

\begin{abstract}
We use numerical simulations to investigate the interaction of an $f$-mode wave packet with small and large models of a sunspot in a stratified atmosphere. While a loose cluster model has been largely studied before, we focus in this study on the scattering from an ensemble of tightly compact tubes. We showed that the small compact cluster produces a slight distorted scattered wave field in the transverse direction, which can be attributed to the simultaneous oscillations of the pairs of tubes within the cluster aligned in a perpendicular direction to the incoming wave. However, no signature of a multiple-scattering regime has been observed from this model, while it has been clearly observable for the large compact cluster model. Furthermore, we pointed out the importance of the geometrical shape of the monolithic model on the interaction of $f$-mode waves with a sunspot in a high frequency range ($\nu=$ 5 mHz). 
These results are a contribution to the observational effort to distinguish seismically between different configurations of magnetic flux tubes within sunspots and plage.
\end{abstract}






\section{Introduction}
Sunspots are an obvious and significant manifestation of a solar magnetic field. An accurate knowledge of the structure of sunspot is essential to understand the magnetic activity of our star.
However, the subsurface structure of these magnetic features is still an unsolved question in solar physics; is it a single monolithic flux tube (monolithic model) as suggested by \citet{cowling53}  or rather a bundle of individual flux tubes like a spaghetti (cluster model) as proposed by \citet{parker79}? 

Observations by \citet{braun87,braun88} showed a significant absorption of $f$-and $p$-modes by sunspots.
In monolithic sunspot model, the absorption is due to the conversion of the incoming acoustic $p$-modes into magnetoacoustic slow $s$-modes that propagate along the magnetic field lines \citep{cally03}. For the cluster sunspot model, 
 the absorption is caused by a multiple-scattering regime from the excitation of tube waves \citep{bogdan91}. A comparison between the scattered wave field of the two competing models shows a remarkable difference, which can be used to distinguish the structure of the model \citep{keppens94}. We have to note that no gravitational stratification was considered in the models studied in the two latter references.

\citet{jain09} used a semi-analytic method to examine the absorption of $p$-modes by a large collection of thin magnetic tubes (plage) in a stratified media. However, they did not take into account the scattering between tubes considering each tube isolated from the others. 
\citet{hanasoge09} were the first to study analytically the multiple-scattering regime of pairs of flux tubes in a stratified atmosphere. They found that the scattering for the kink mode ($m=\pm1$) changes dramatically for small flux-tubes separations. They showed also a significant contribution of the near-field phenomenon on the scattering. \citet{hindman12} used scattering formalism to investigate the interaction of the monopole component ($m=0$) of acoustic $p$-modes with a thin magnetic fibril. They obtained that mode-mixing and absorption are weak for thin flux tubes.
\citet{hanson14a} used the semi-analytical model of \citet{hanasoge09} to incorporate the sausage mode in addition to the kink one, showing the importance contribution of the sausage mode on the scattering by the pair of tubes.
They concluded that the absorption of sausage mode is a magnitude larger than that of the kink mode when the tubes are in a close proximity for the higher frequency of 5 mHz. \citet{hanson14b} extended the model of \citet{hanson14a} to study the scattering by a larger ensemble of thin magnetic flux tubes. They deducted that the absorption enhanced for a larger ensemble of tubes, or higher frequency. In addition, they noted that the spatial distribution of tubes affects the absorption at higher frequencies (5 mHz).

Numerical simulations of wave propagation through solar magnetic features provide an efficient and direct way to infer their structure
by observing their helioseismic signatures. Recently, \citet{felipe13} investigated numerically the interaction of $f$-mode with an ensemble of flux tubes of different number and configuration. They found that the multiple-scattering affects strongly the absorption coefficients,  showing that the sausage and kink modes are the dominant modes for the scattering. They noted also that the absorption generally increases with the number of flux tubes and the reduction of the distance between them. 
\citet{felipe14} studied the helioseismic signatures of monolithic and spaghetti sunspot models, where the latter model contains a realistic number of tubes for the first time. They obtained that the mode-mixing from the monolithic model is more efficient than that of the spaghetti model. Their simulations reveal also that the differences observed in the absorption coefficient for both models can be detected above the noise level.

\begin{figure}
 \includegraphics[width=8cm]{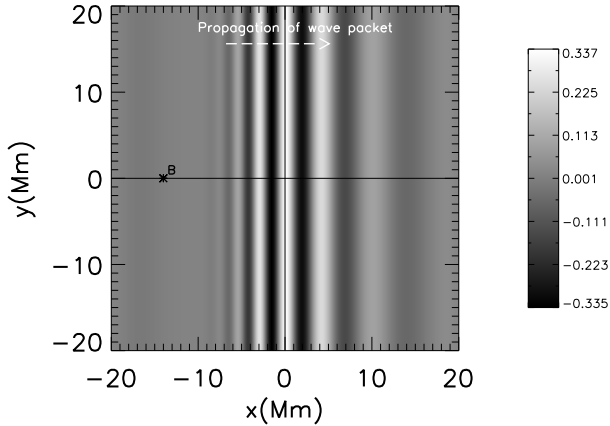} 
\vspace{0.03\textwidth}
 \caption{Snapshot of the full unperturbed wave field of $V_z$ at $t=3300$ s.}
   \label{fig1}
\end{figure}

\citet{daiffallah14} simulated the propagation of an $f$-mode wave packet centered at 3 mHz frequency through a hexagonal cluster of thin magnetic flux tubes, considering the effect of tubes configuration and the separation distance on the scattering. It was found that when the separation between two neighboring tubes within the cluster $d$ is about $\lambda/2\pi$ ($\lambda$ is the wavelength of the incoming wave), individual tubes within the loose cluster start to scatter waves to nearby tubes which scatter again to a near field and so on leading to a greatly enhanced absorption measured in the far field (multiple-scattering regime). We define a compact cluster as a bundle of magnetic flux tubes in a close-packed configuration.
We have shown that a loose cluster in a multiple-scattering regime is a more efficient absorber of waves than a compact cluster or an equivalent monolithic tube of both clusters. 

In the present study, unlike \citet{daiffallah14}, we have fixed the distance $d$ as in compact cluster separation, and we have changed the wavelength $\lambda$ of the incoming wave to see if the multiple-scattering regime can occurs for a cluster in a close-packed configuration as in the case of a loose cluster in the same regime. Given this, an important question emerges: is the condition on $d$ (cited above) sufficient to have a multiple-scattering regime for the compact cluster or we have to add the condition that the cluster must be in a loose configuration to allow tubes to communicate through their near field. An important part of this work will try to answer this question.

The paper is organized as follows. In section \ref{S-simulations} we briefly describe the code that we used and the set up of the simulations. In Section \ref{S-identmsr} we present the method of inspecting the multiple-scattering effects. Sections \ref{S-smallspots} and \ref{S-largespots} outline the results of the interaction of a wave with small and large size models of sunspot respectively in a stratified atmosphere, including the effect of the frequency variation on the scattering for both cases. For more general results, we compute in section \ref{scatc} the scattering cross section for the different sunspot models. Finally, the discussion and conclusions are presented in section \ref{concl}.


\section{Simulations}
     \label{S-simulations} 
We have performed the simulations using the {\sf{SLiM}} code \citep{cameron07} which solves the linear and ideal MHD equations in a three-dimensional stratified atmosphere. A pseudo-spectral scheme is implemented in the horizontal directions and a two-steps Lax-Wendroff scheme in the vertical direction to evolve the horizontal Fourier modes. The background atmosphere is an enhanced polytropic atmosphere described by \citet{cally97}. The horizontal extent of the computational domain is $x$ $\in [-20,20]$ Mm and $y$ $\in [-20,20]$ Mm. The depth $z$ is ranged from 0.2 Mm to 6 Mm below the solar surface.
The spatial resolution for all simulations is 192 $\times$ 192 Fourier modes in $x$-and $y$-directions, where it is 150 grid points in the $z$ direction.
 
Since $f$-mode interacts strongly with the sunspot compared to the $p$-mode \citep{bogdan96,zhao13}, we propagate in all our simulations an $f$-mode wave packet with a Gaussian envelope centered at the angular frequency $\nu = 3$ mHz with a standard deviation of 1.18 mHz. Each individual $f$-mode has an exponential dependance on depth $z$ with the frequency $\omega=(gk)^{1/2}$ where $k$ is the horizontal wavenumber.

In subsection \ref{S-fsmallspots} and \ref{S-flargespots} we have studied the scattering using centered frequencies from $\nu = 2$ mHz  to $\nu = 5$ mHz. 

At $t = t_0$, the
wave packet is situated at the left edge of the computational domain $x_0 = -20$ Mm and it propagates from the left to the right in the $x$-direction.   
All waves of the $f$-mode are in phase at the initial position.

\begin{figure*}
\center
 \includegraphics[width=5cm]{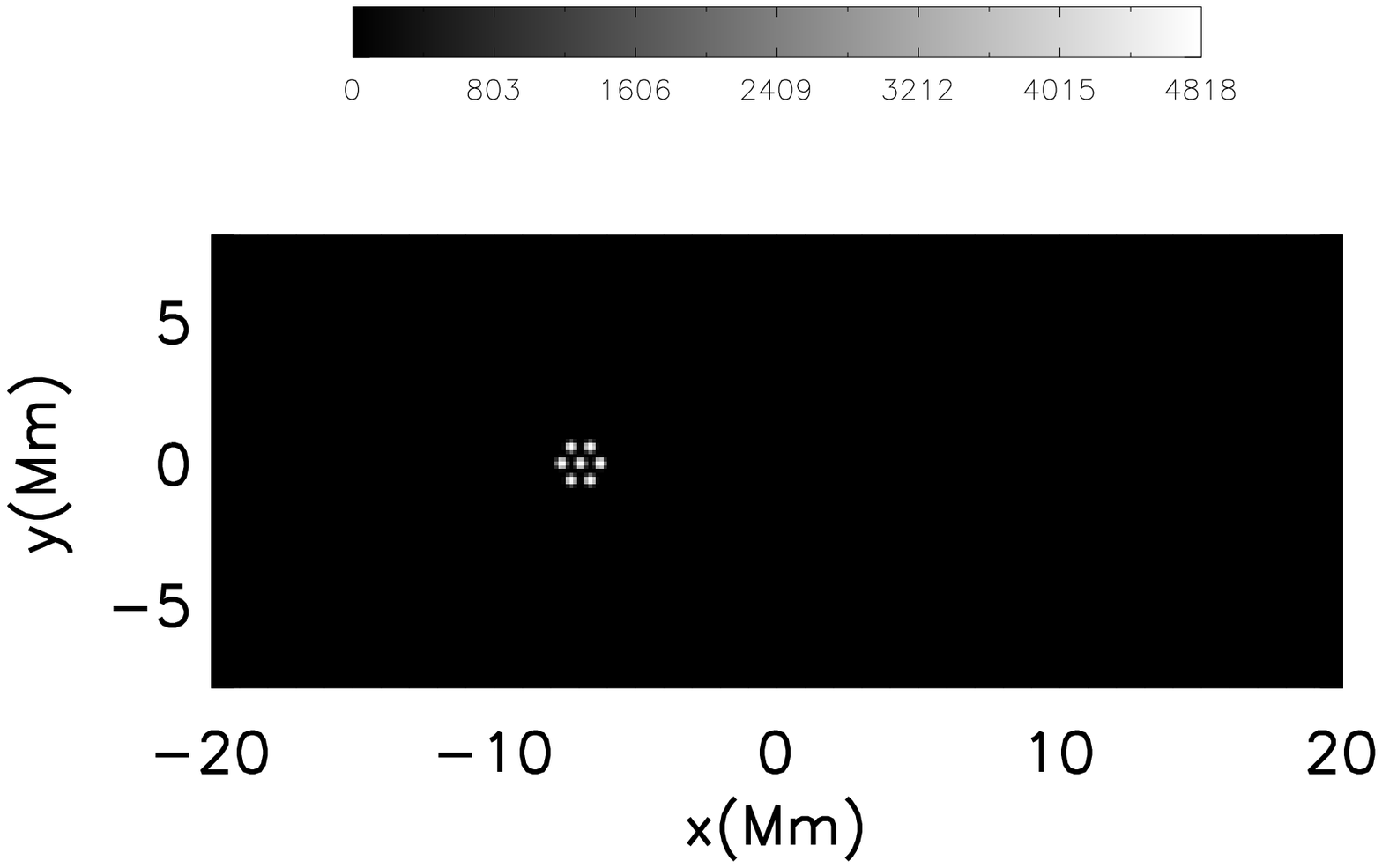} 
\includegraphics[width=9cm]{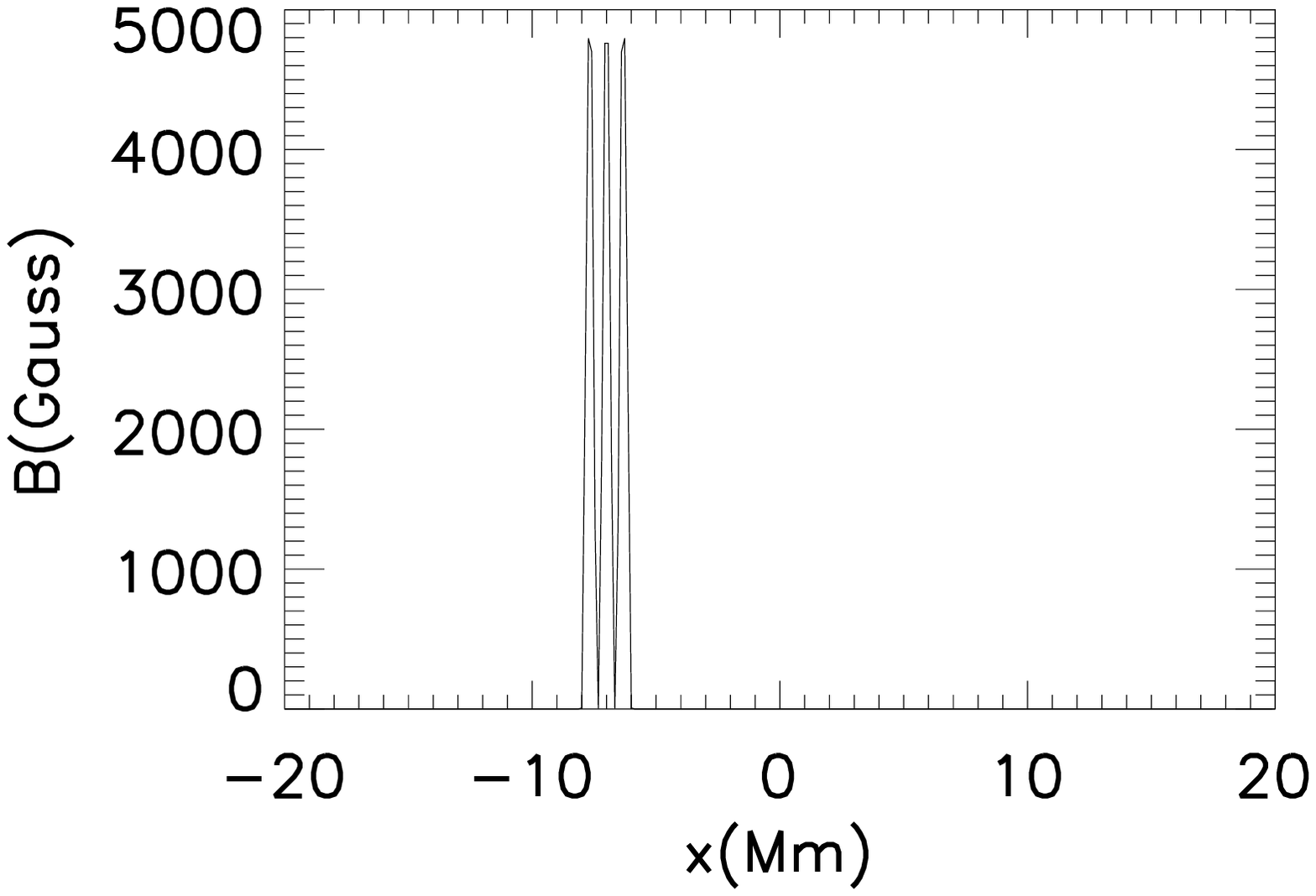} 
\includegraphics[width=5cm]{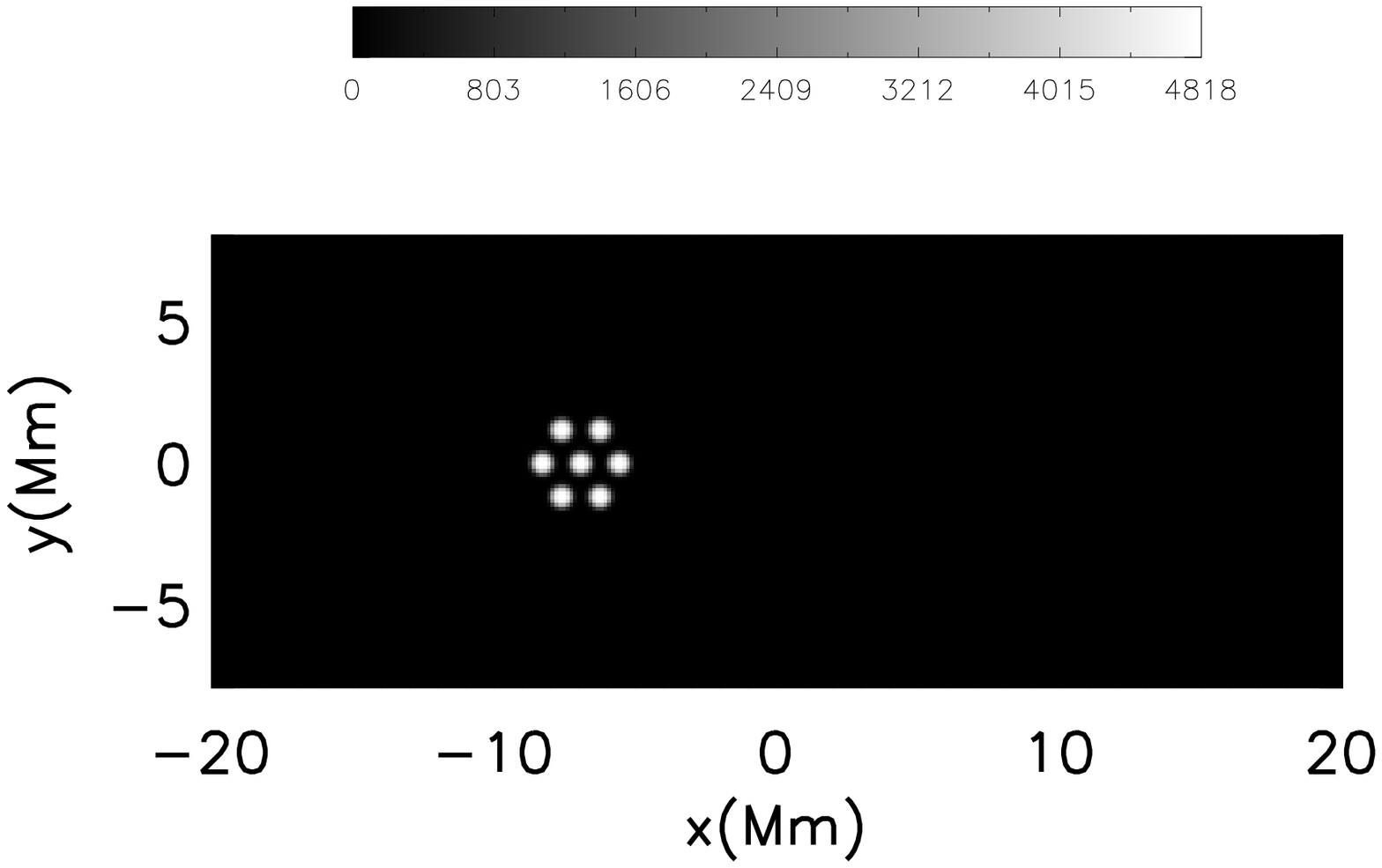} 
\includegraphics[width=9cm]{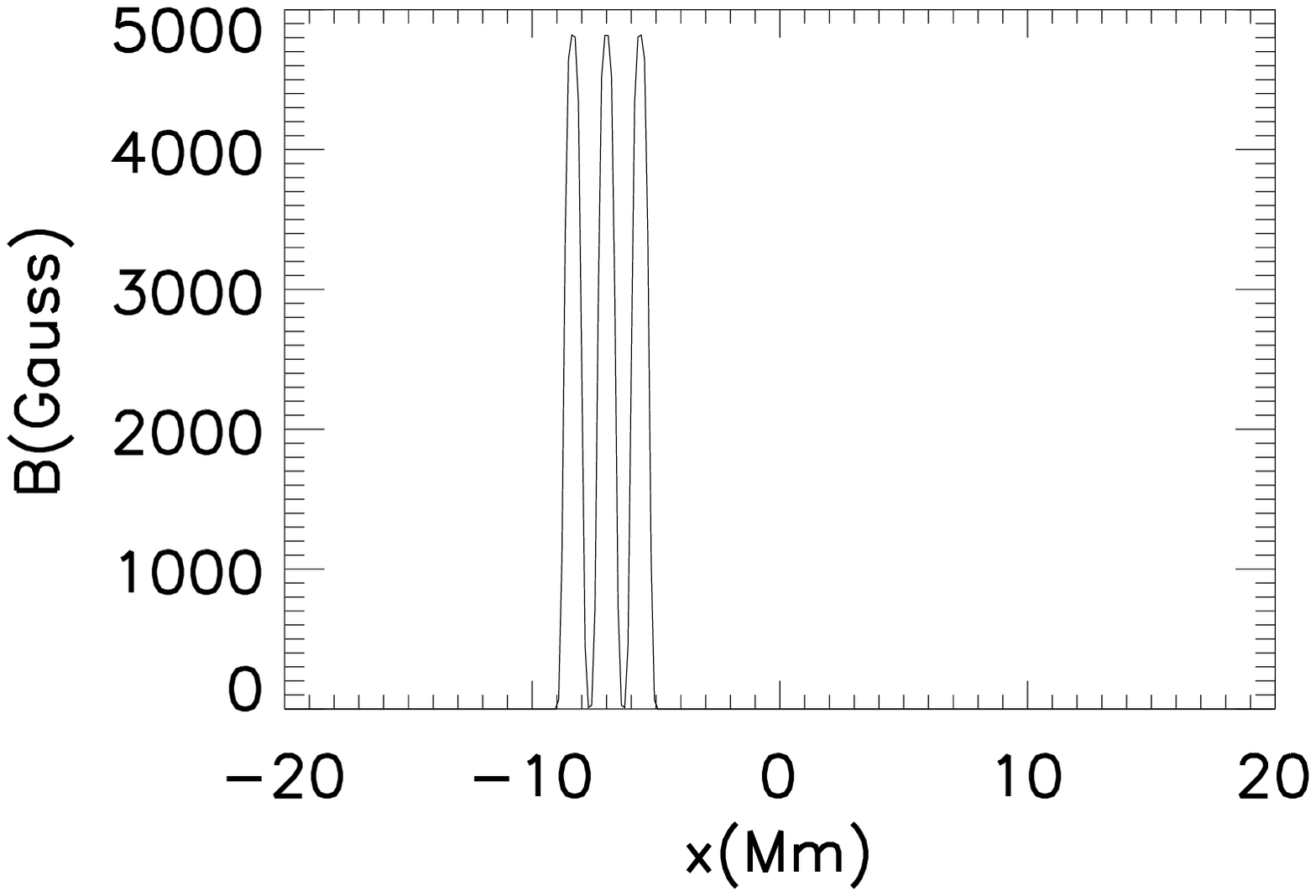} 
 \caption{The magnetic field profile from the simulated data. We define $R_c$ as the radius of an  individual tube within the cluster. The top panel corresponds to the small cluster ($R_c$ = 200 km) where the bottom panel is for the large cluster ($R_c$ = 400 km).  The snapshots show the magnetic field strength taken at the surface ($z$ = 280 km) in $x-y$ plane. The curves show the magnetic field profile along the line $y=0$. The separation distance between two neighboring tubes for the small and the large compact clusters is fixed as  $d=3.4R_c$.}
   \label{fig2}
\end{figure*}

A periodic boundary condition is imposed on the horizontal 
side walls of the simulation box. 

The upper condition plays an important role in the absorption and the scattering of $f$-modes. In our case, they correspond to that of a free-surface in the non-magnetic regions where waves are strongly reflected from the upper boundary \citep{cally97,cameron08}. Our initial condition has a very little energy in the range where it would escape ($<$ 0.06 \% ). Recently, \citet{jain14} have modeled the interaction of $f$-and-$p$-modes with a random distribution of tubes. They have found that the absorption coefficients and the damping rates are very sensitive when a magnetic fibril is extended into an isothermal region above a polytrope. More recently, \citet{hanson15} have extended the model of \citet{hanson14b} by allowing sausage and kink modes to freely escape at the top of model using a radiative boundary condition there. Their results show an increase of the absorption coefficiens of the incoming $f$-wave for both modes in comparison to the absorption using a reflective stress-free condition at the top of tubes.
Our condition is not realistic since in the realistic one, waves can escape to the chromosphere and corona. However, our goal is not to construct a model that mimics the solar atmosphere, but to study a very simple sunspot models where we have control over the physics.

At the bottom layer, we impose that waves are evanescent.

The initial individual 
magnetic flux tube is vertical in the $z$-direction. It is embedded
in the polytropic background  atmosphere  with a radial top profile given by  $ B(r) = B_{0} \exp(- r^4/ R^4) $ where $R$ 
is the tube radius, and $B_{0}$ = 4820 G \citep{cally97}. The magnetic field has the same radial profile along the depth $z$. The center of the sunspot model is located at the point $x=-7$ Mm, $y = 0$. 

The simulation without the flux tube corresponds to $B_0 = 0$. The scattered wave field is obtained by subtracting the simulation without the flux tube from the simulation with the flux tube.

We consider that the vertical velocity $V_z$ at the upper surface is the most appropriate component to analyze the scattering since it is the only component that can be measured with Dopplergrams from the solar disk center. All the slices of the scattered vertical velocity in this paper were taken near the surface at the depth $z=280$ km. Figure \ref{fig1} shows the unperturbed full wave field of the component $V_z$ taken at $t=3300$ seconds. The amplitude of the incoming wave is normalized at 1. This will set the scale for the scattered wave.

According to the variation of the sound and the Alfv\'en speeds in the enhanced polytropic atmosphere, the plasma-$\beta$ is not constant inside the tube. These speeds are set to be equal at a depth of 400 km where the plasma-$\beta \approx 1$.

In this paper, we consider only the $f$-$f$ scattering without focusing on the phase shift variation, where $f$-$p_n$ mode-mixing decreases rapidly with increasing radial number $n$ \citep{hanasoge08,felipe12,zhao13}.

In our simulations, the monolithic models with a large radius can be considered as thick tubes with respect to the wavelength of the incoming wave. The coupling between the fast and the slow magnetoacoustic waves makes the distinction between different modes difficult, particularly near the surface where $\beta=1$. Consequently, these tubes will scatter in all $m$ and the vertical velocity will appear as a summation of all these modes.

\begin{figure} 
\centering 
\includegraphics[width=8.5cm]{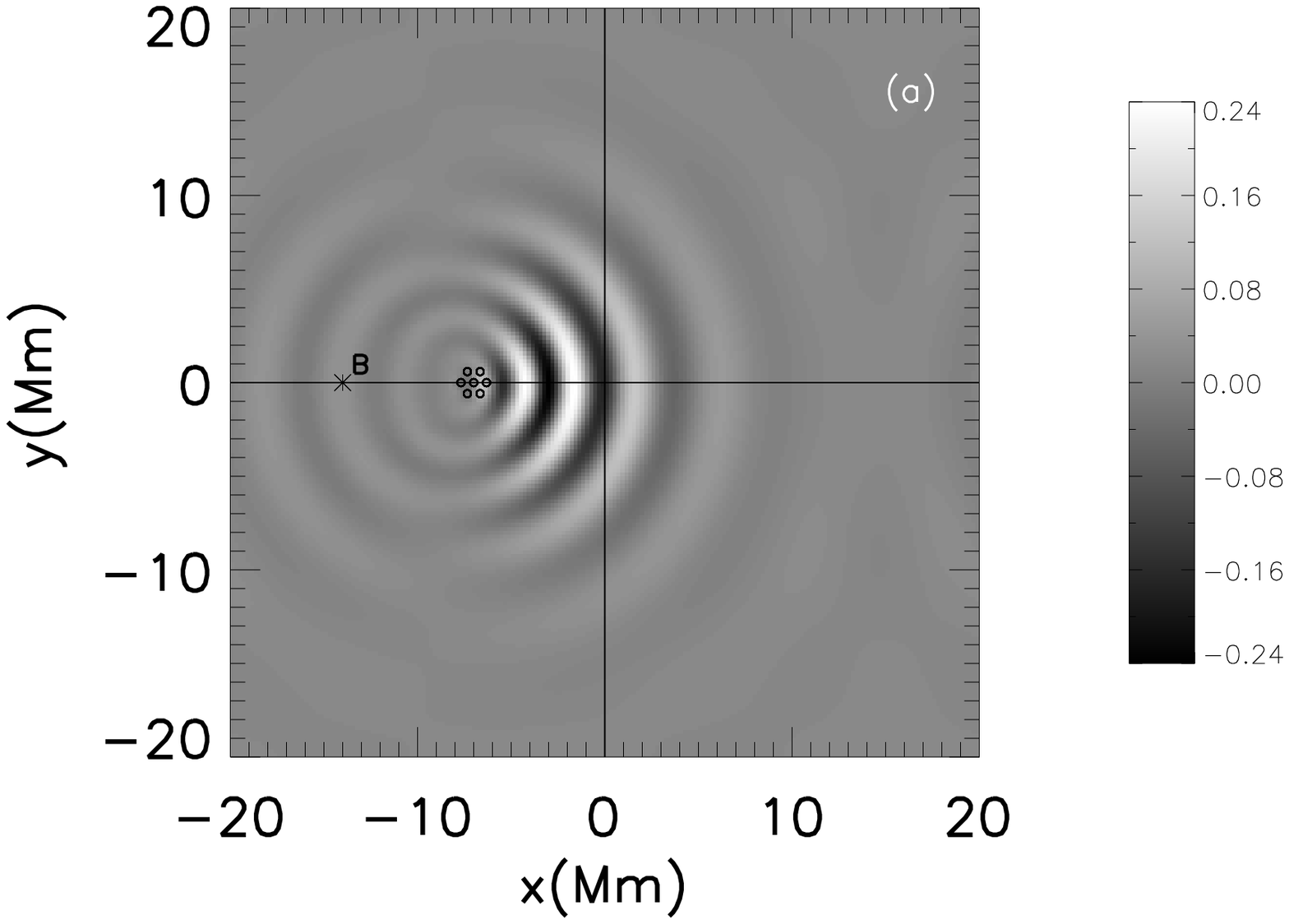} 
\includegraphics[width=8.5cm]{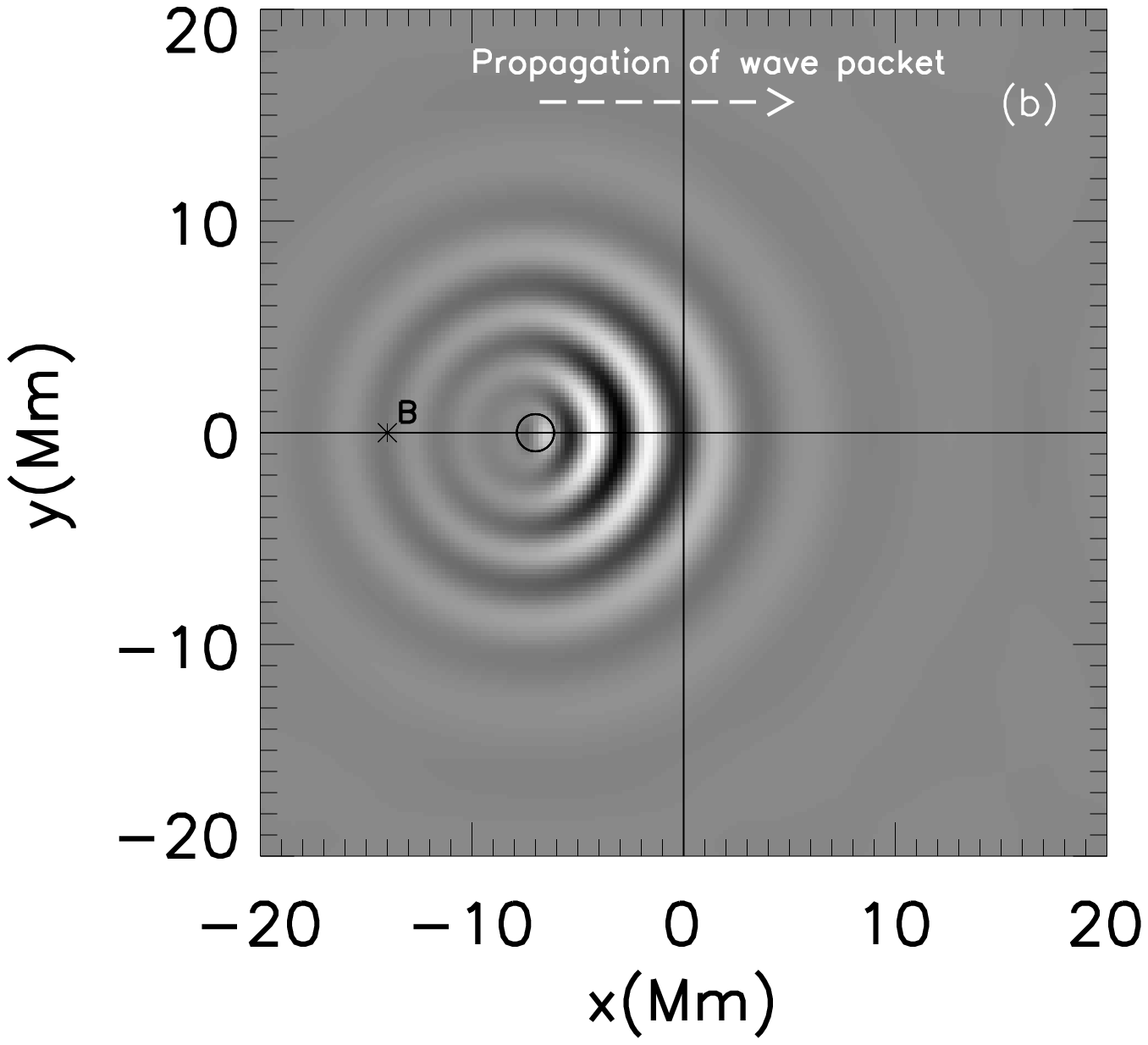} 
\includegraphics[width=8.5cm]{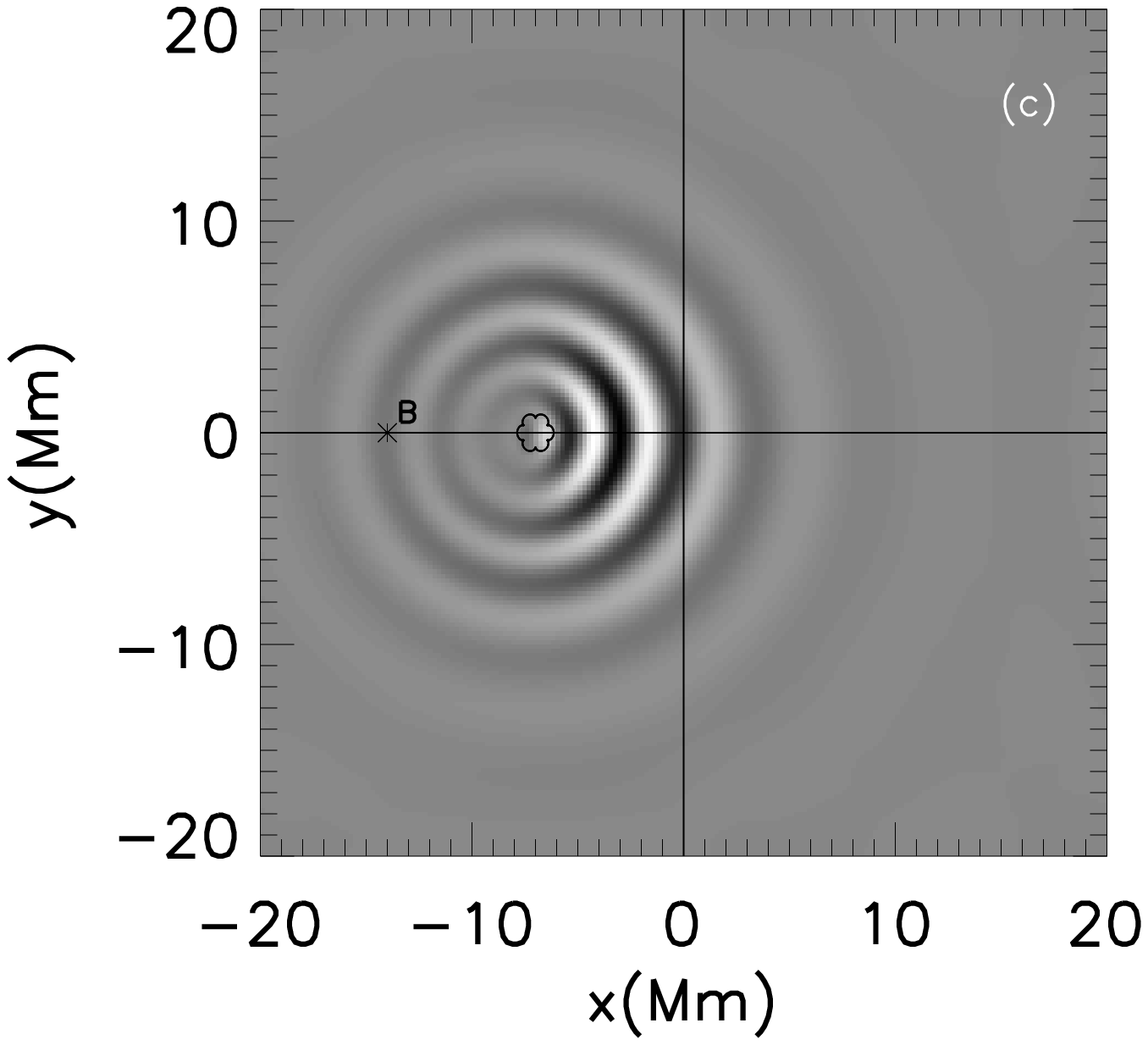} 
\vspace{0.03\textwidth}
 \caption {Scattered wave field of $V_z$ at $t$ = 3300 s for 
three different models of sunspot ($\nu=3$ mHz): (a) a cluster of seven identical compact tubes of 200 km radius, (b) a single monolithic tube whose radius $R=880$ km is an average radius of the cluster, (c) a hexagonal monolithic tube where the size at the surface is the same as that of the cluster. The colour scale is the same for the three snapshots. The scattered wave fields of the monolithic tube (b) and its equivalent hexagonal monolithic model (c)  seem to be similar, whereas the compact cluster model (a) shows a more extensive left wave field in the $y$-direction.} 
\label{fig3}
\end{figure}


\section{The Identification Method of the Multiple-Scattering Regime}
\label{S-identmsr} 

We are inspecting the multiple-scattering from two complementary points of view:

1- From a visual inspection of the scattered wave field; we compare the scattered wave field of a monolithic model with that of cluster models. As the $f$-mode has a maximum of power at the surface, multiple-scattering from a cluster model is easily identifiable when individual tubes scatter waves to the near field making their specific signatures in that region. In the Figure 9 of \citet{daiffallah14}, we have observed a multiple-scattering signature from two loose clusters made of 7 and 9 tubes respectively ($d=0.2 \lambda$). The near-field area extends till a distance of 2 Mm from these structures in the $x$-direction and it can be clearly distinguished from the wave field of the equivalent monolithic tube.        \\ 

2- From the temporal profiles of the scattered surface vertical velocity measured at a single point situated in the far field. Actually, the multiple-scattering regime enhances the absorption of the incoming wave (near-field scattering) leading to a decrease in the amplitude of the scattering measured in the far field. The visual inspection of multiple-scattering effects from a pair of magnetic flux tubes and from two loose clusters made of 7 and 9 tubes was confirmed by using this method \citep{daiffallah14}.

We have shown  in \citet{daiffallah14} that the multiple-scattering regime occurs for a separation distance  $0.12 \lambda < d \le 0.2 \lambda$  which is approximately $d \sim \lambda/2\pi$ in the case of 3 mHz $f$-mode.
However, for a separation  $0.2\lambda < d \lesssim \lambda/2$, we have got a coherent scattering regime which is characterized by the enhancement of the scattering in both near field and far field, but no absorption was measured in the far field. Since both ranges correspond to the scattering regime, we can merge them to get  $ \lambda/2\pi \le d \le \lambda/2$. In conclusion, we can state that the scattering regime occurs when the separation distance between tubes within cluster is  $\lambda/2\pi \le d \le \lambda/2$, where $d \sim \lambda/2\pi$ is the lower limit and $d\sim \lambda/2$ is the upper limit. The latter value is consistent with previous studies by \citet{hanasoge09} and \citet{felipe13} who agree that the extent of the region of influence of the near field in multiple-scattering is $\lambda$/2.

\begin{figure*}  
\centering
\includegraphics[width=0.35\textwidth]{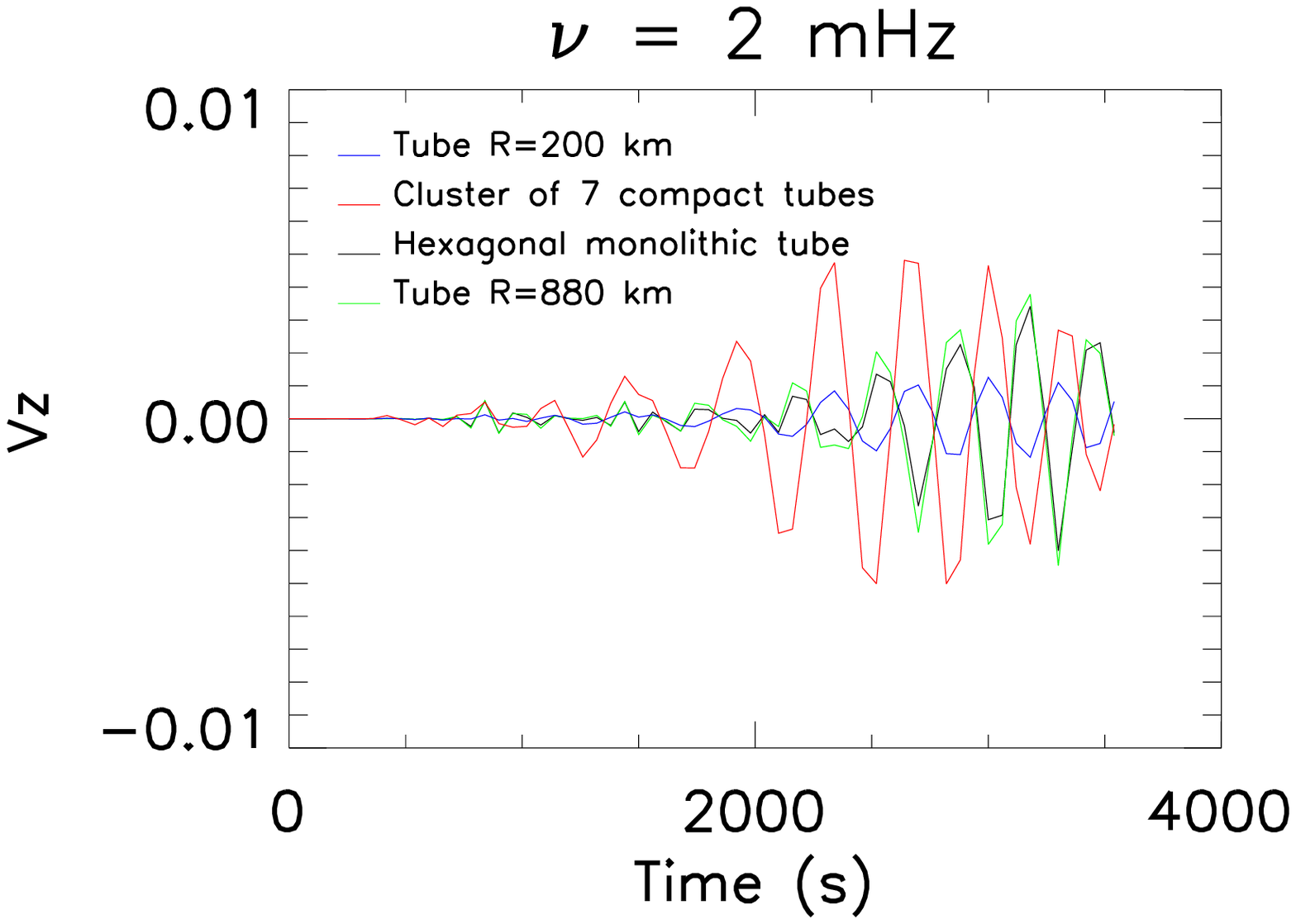} 
\includegraphics[width=0.35\textwidth]{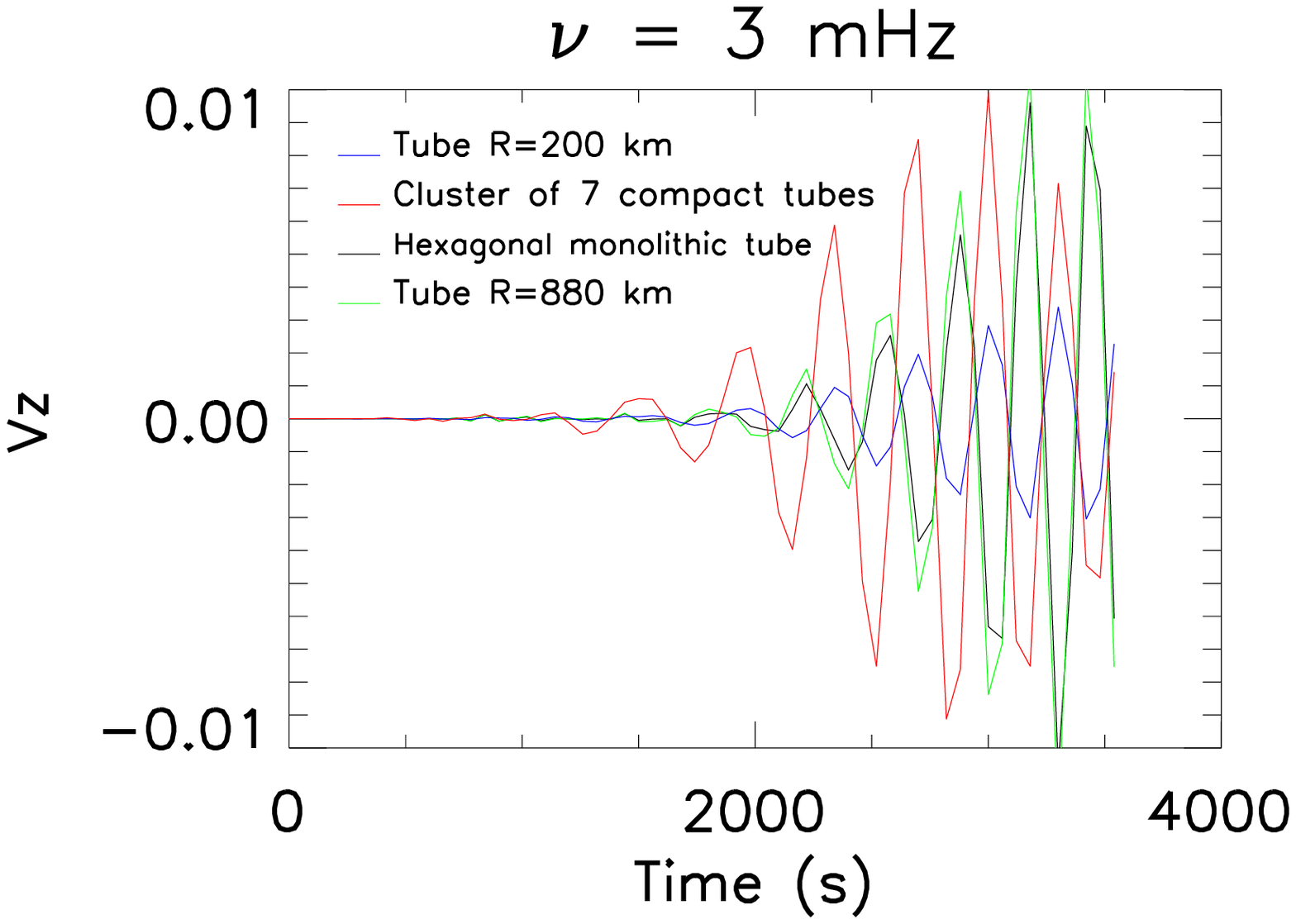}
\includegraphics[width=0.35\textwidth]{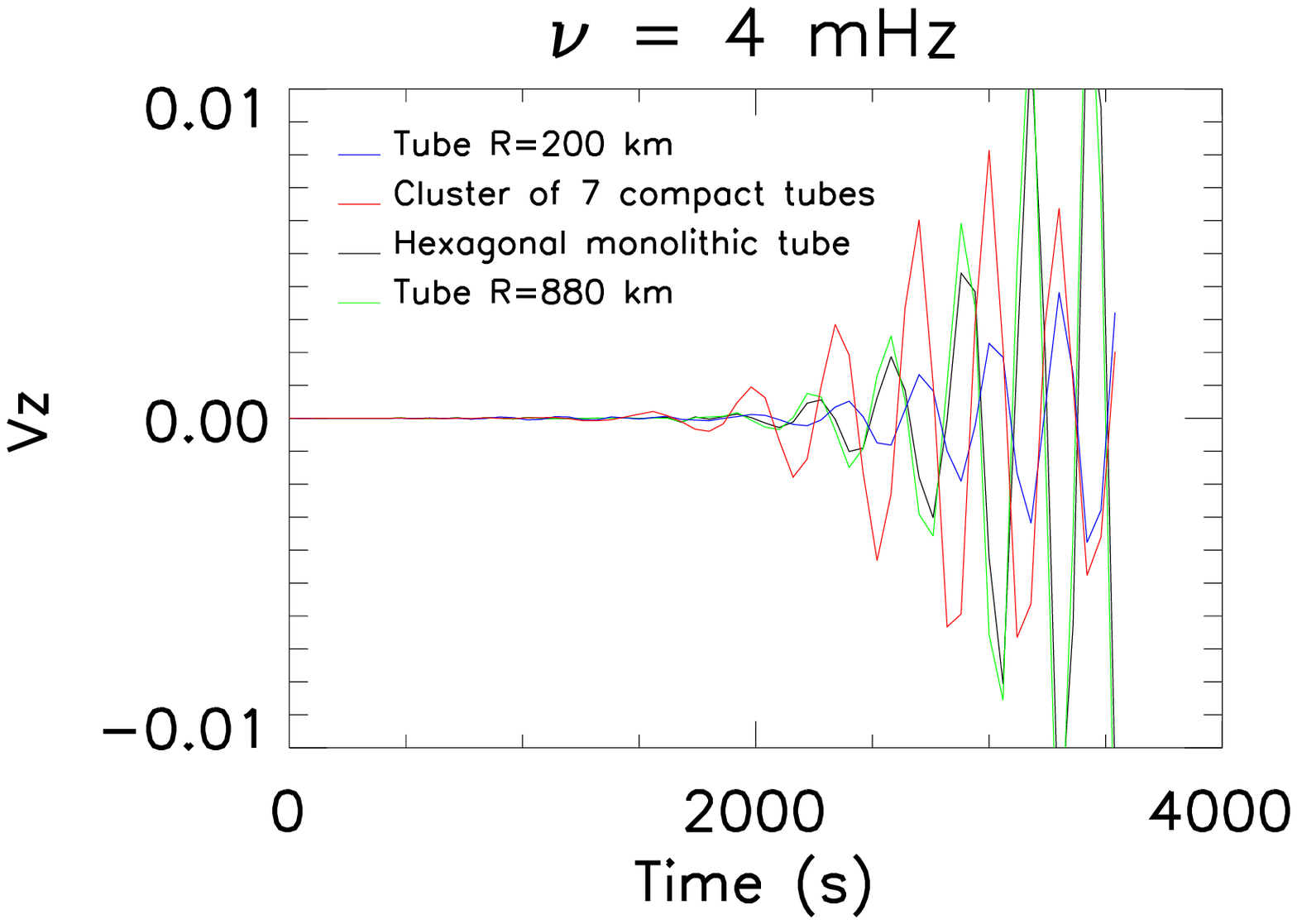}
\includegraphics[width=0.35\textwidth]{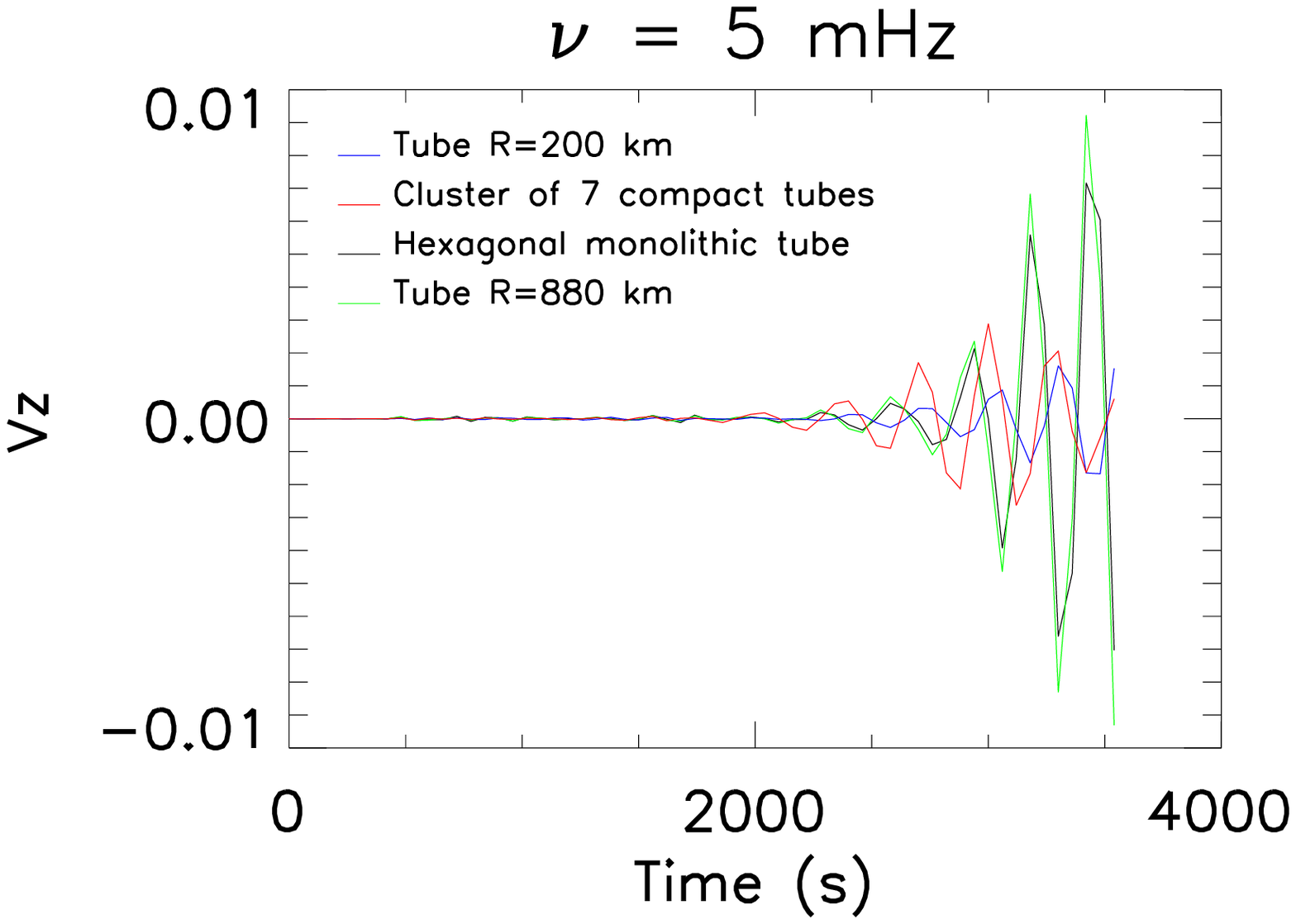}

 \caption {Scattered vertical velocity $V_z$ as a function of time measured at  point B for the different magnetic features (a), (b), (c) and for a single monolithic tube of 200 km radius. The curves are plotted for the $f$-mode frequencies 2 mHz, 3 mHz, 4 mHz, and 5 mHz.} 
\label{fig4}
\end{figure*}


\section{$f$-mode Interaction with Small Sunspot Models}
     \label{S-smallspots} 

To probe the subsurface structure of sunspots, it is essential to construct adequate sunspot models that mimic the full complexity of solar magnetic structures. However, our aim is not to study a realistic sunspot to reproduce quantitatively the observations, but to get specific seismic signatures from simple sunspot models with basic properties in order to distinguish between them, at least qualitatively. 

In addition to monolithic and cluster models of the sunspot that were studied by \citet{daiffallah14} and previous papers, we have incorporated here an intermediate new model to better interpret and understand the results.

We have carried out three simulations shown in Figure \ref{fig3} (a,b,c) where the scattered wave field at $t$ = 3300 s is displayed.  
The snapshots show the propagation of an $f$-mode wave packet of angular frequency $\nu =$ 3 mHz through:

\begin{itemize}
 \item (a) a cluster of seven identical flux tubes in hexagonal compact configuration. Each individual tube within the cluster has a radius $R_c=200$ km,
 \item (b) a single monolithic tube whose radius $R$ = 880 km is the average radius of the cluster in (a), 
\item (c) a hexagonal monolithic flux tube in which the shape at the surface is the same as that of the cluster in (a).
\end{itemize}

Figure \ref{fig2} to the top shows the magnetic field profile of the cluster model (a) from the simulated data. The separation distance between two neighboring tubes within the compact clusters is fixed as  $d=3.4R_c$, where $R_c$ is the radius of an individual tube within the cluster.

\begin{figure} 
\includegraphics[width=8.5cm]{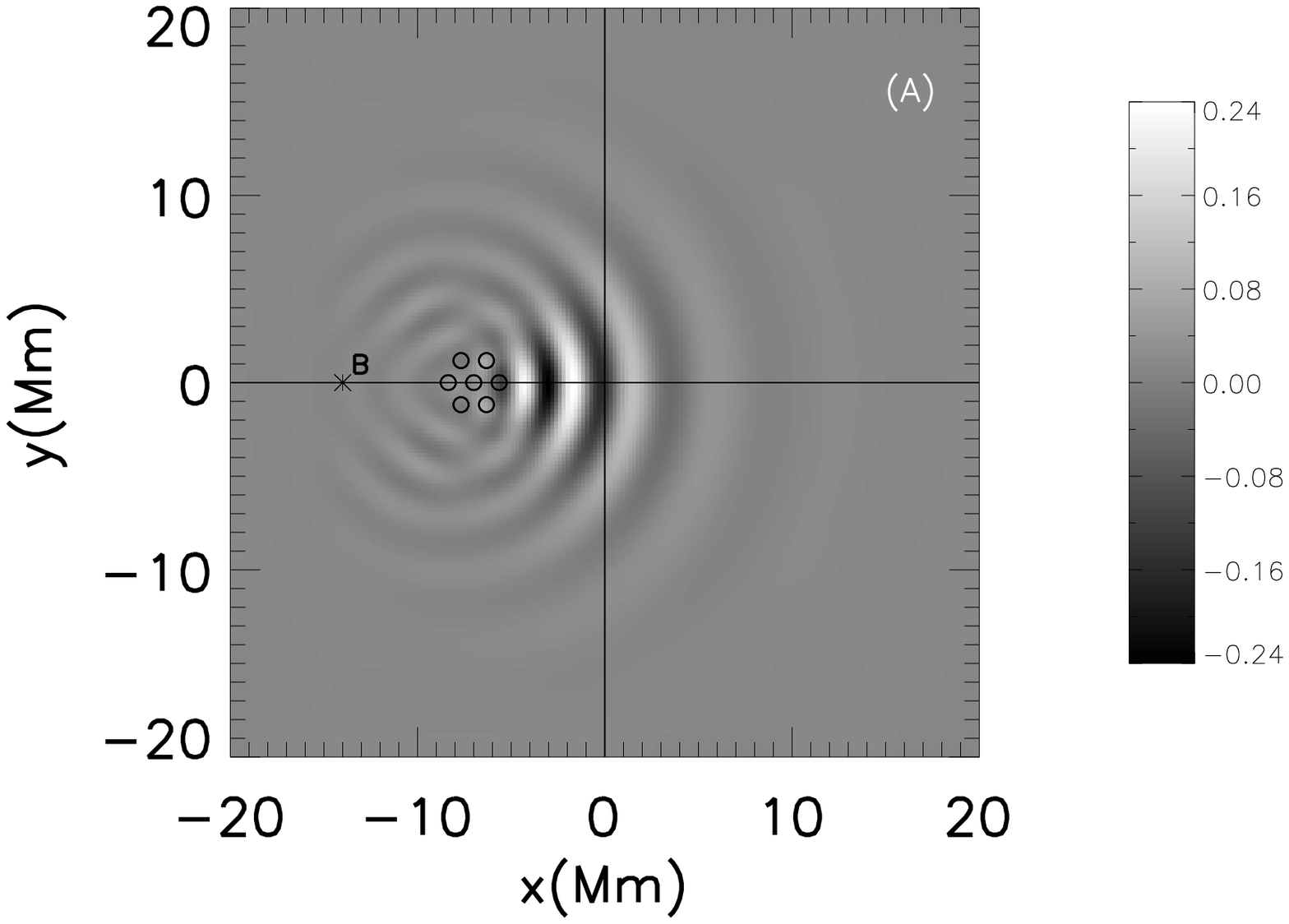} 
\includegraphics[width=8.5cm]{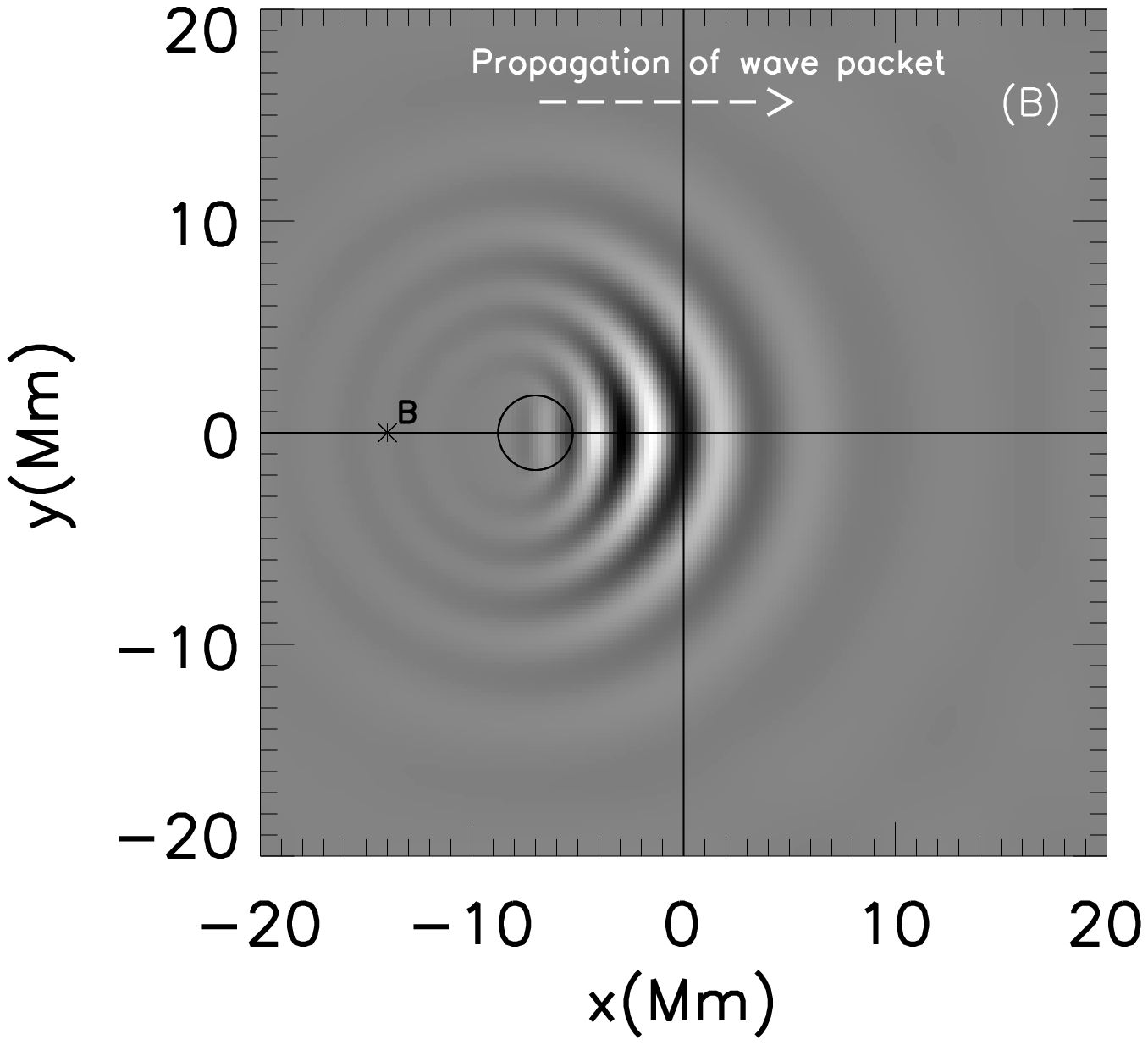} 
\includegraphics[width=8.5cm]{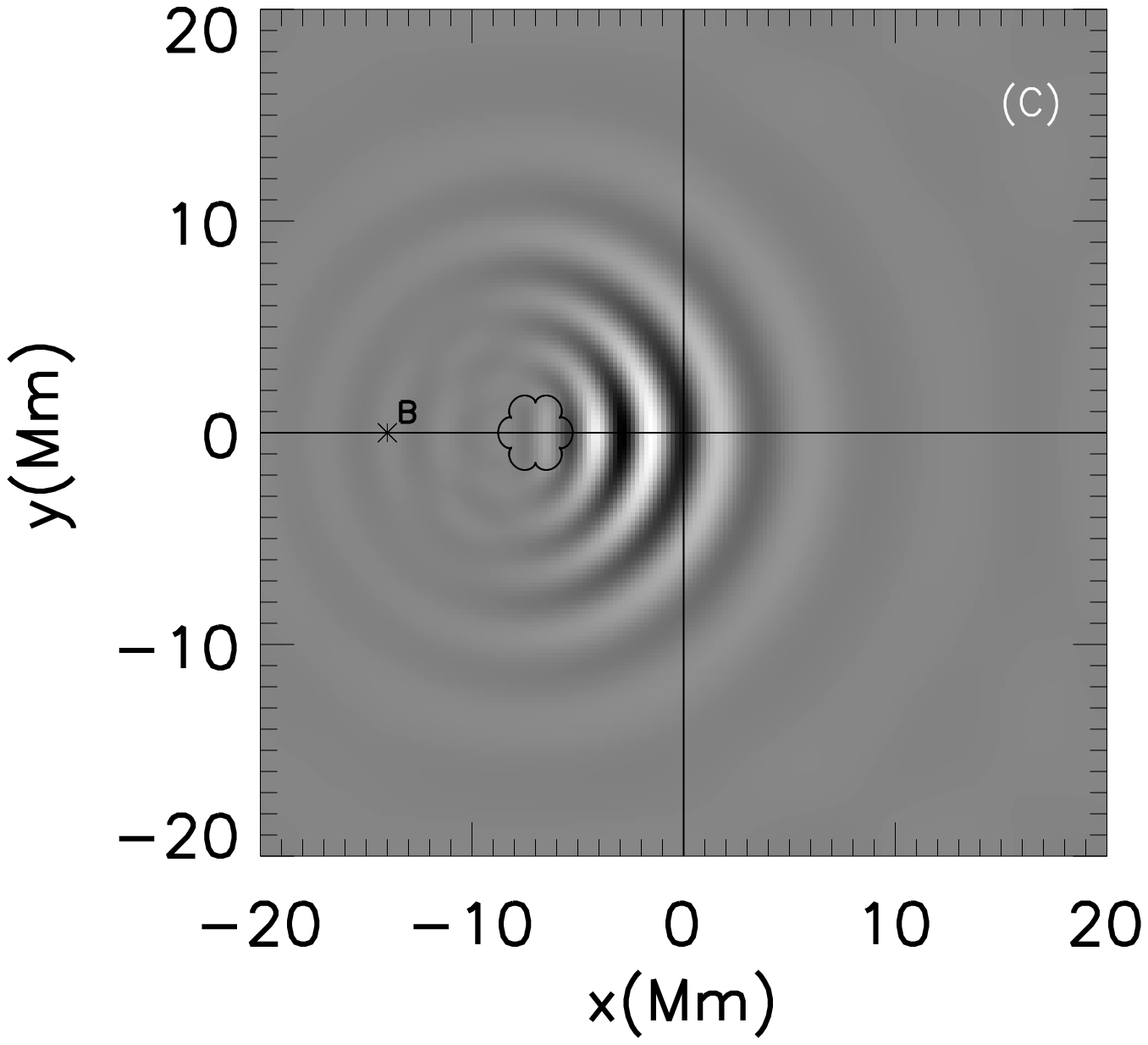} 
\vspace{0.03\textwidth}
 \caption {Scattered wave field of $V_z$ at $t$ = 3300 s for 
three different models of sunspot ($\nu=$ 3 mHz): (A) a cluster of seven identical compact tubes of 400 km radius, (B) a single monolithic tube whose radius $R=1.76$ Mm is an average radius of the cluster, (C) a hexagonal monolithic tube where the size at the surface is the same as that of the cluster. The colour scale is the same for the three snapshots. The compact cluster model (A) shows a triangular shape waveform in the left near field, whereas the hexagonal monolithic model (C) shows a non-uniform waveform.}
\label{fig5}
\end{figure}

The model (c) is an intermediate feature between (a) and (b). It is interesting to see what can be the contribution of the sunspot on the scattering 
in terms of geometrical shape. We note that the contour of the magnetic field in hexagonal model (c) is obtained by using the parametric equation of an epicycloid ($q=6$).

We recall that we will restrict our analysis on the left scattered wave field to the magnetic elements where their oscillations are observed without a contribution from the incoming wave.
Figure \ref{fig3}(a) shows the scattering from the cluster. No signature of multiple-scattering regime was observed from this model. However, we observe that waves seem to be slightly compressed in the $x$-direction compared to the circular waveform of the monolithic tube of 880 km radius. In fact, \citet{daiffallah14} showed that a pair of flux tubes aligned perpendicular to the direction of the incoming wave oscillate simultaneously with the $f$-mode in $y$-direction whatever the separation $d$ is. Indeed, we can distinguish within the cluster two pairs of tubes in a perpendicular configuration to the incoming wave.

\citet{felipe14} observed in their simulations that waves ($f$-mode) scattered by a spaghetti model of the sunspot are more flat compared to that scattered by a monolithic tube. It is reasonable to infer that the  flattening observed in our simulation is the same phenomenon described by these authors. The oscillation of tubes in $y$-direction when wave propagates contributes to the scattered wave giving this appearance. This effect could be amplified if there are more tubes inside the bundle. 
  
As is apparent from Figure \ref{fig3}, the scattered wave fields of the hexagonal monolithic model (c) and its equivalent monolithic tube (b) seem to be similar, at least qualitatively.


\subsection{Multi-Frequency Effects on the Scattering}
     \label{S-fsmallspots} 

To complete this study, we need to investigate how oscillations from sunspot models vary with the frequency. To do so, we have fixed the separation distance $d$ as in the cluster (a), and we have changed the wavelength $\lambda$ of the incoming wave through the change in frequency. This will allow us to change the ratio $d/\lambda$ without changing the size of the cluster.
We have studied the scattered wave field for the different models (a), (b) and (c) using four frequencies of $f$-mode. Figure \ref{fig4} shows the scattered vertical velocity $V_z$ as a function of time measured at point B(-14,0) for the different magnetic features (Figure \ref{fig3}) and for four frequencies of $f$-mode (from 2 to 5 mHz).  The amplitude of the incoming wave in an unperturbed field is normalized at 1 (Figure \ref{fig1}).

Firstly, we observe that the curves are more extended in time for $\nu=2$ mHz than for $\nu=5$ mHz where the wave packets are more compressed. This effect shows the variation of the wavelength of the wave packet with the frequency.
As shown in Figure \ref{fig3}, Figure \ref{fig4} confirms that the monolithic tube (b) and the hexagonal monolithic model (c) have the same behaviour and consequently oscillate in phase, while the hexagonal monolithic model shows a reduced amplitude relative to that of the monolithic tube for all frequencies.

The behaviour of the scattered curve from the compact cluster model 
is more interesting. For $\nu=2, 3$ mHz ($d\le\lambda/2\pi$), the curve of the cluster measured in B shows a similar trend as the curve of the  single tube of 200 km radius. At these frequencies, the cluster should oscillate as a monolithic tube of the same size. However, it oscillates with a different mode due to its fibril structure, which demonstrates the particularity of this model with a behaviour completely different from that of the monolithic tube at this frequency.

For $\nu=5$ mHz, the curve of the cluster starts to be out of phase with that of the single tube of 200 km radius. This case corresponds to the multiple-scattering regime in principle since the condition $\lambda/2\pi <d< \lambda/2$ is satisfied. However, no signature of the tubes'waves from the cluster was observed in the near field at this frequency. In fact, individual tubes within the cluster start to oscillate more efficiently in $y$-direction, which explains probably the reduction of the scattering amplitude of the cluster at this frequency.

\begin{figure*}  
\centering
\includegraphics[width=0.35\textwidth]{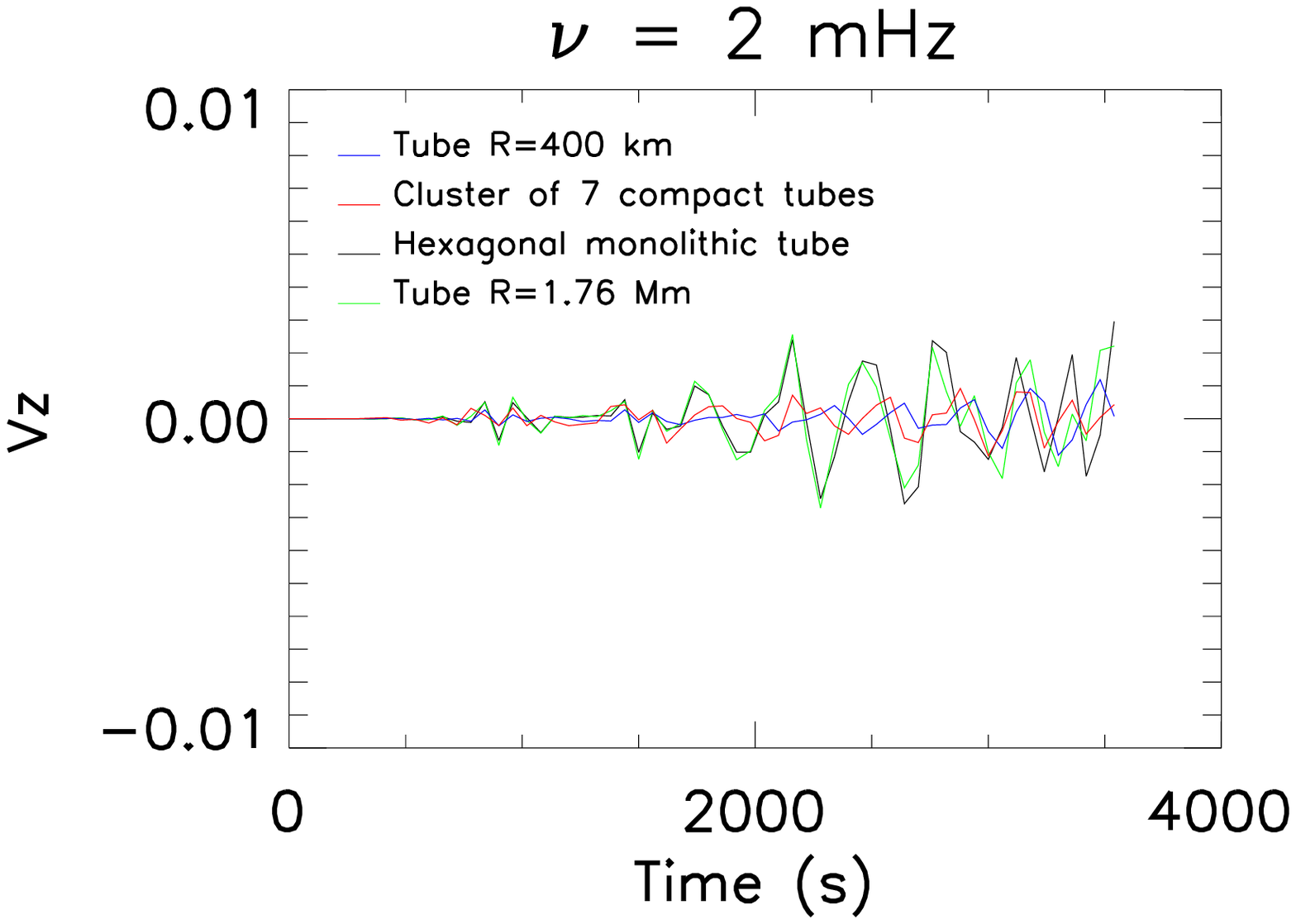} 
\includegraphics[width=0.35\textwidth]{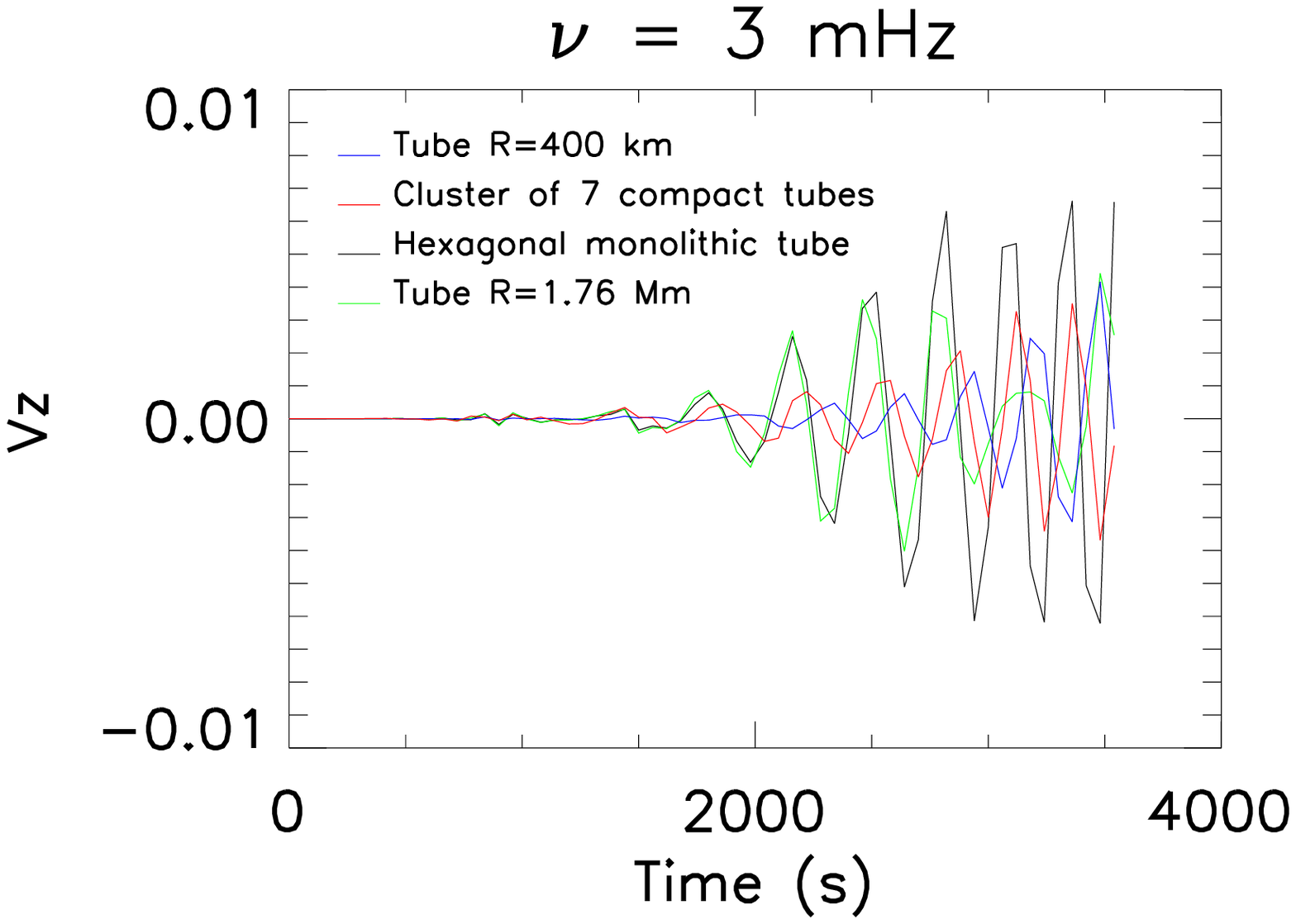}
\includegraphics[width=0.35\textwidth]{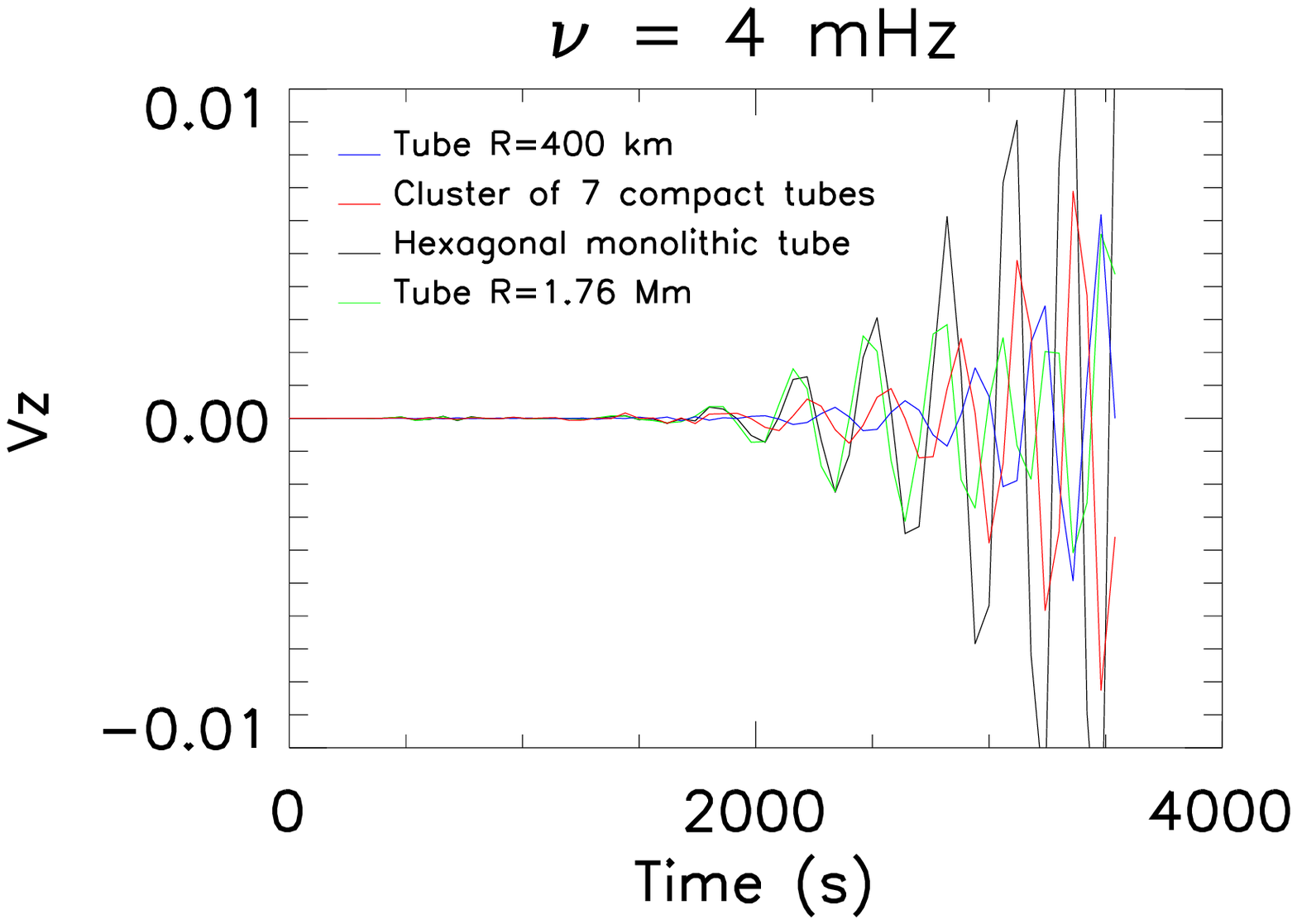}
\includegraphics[width=0.35\textwidth]{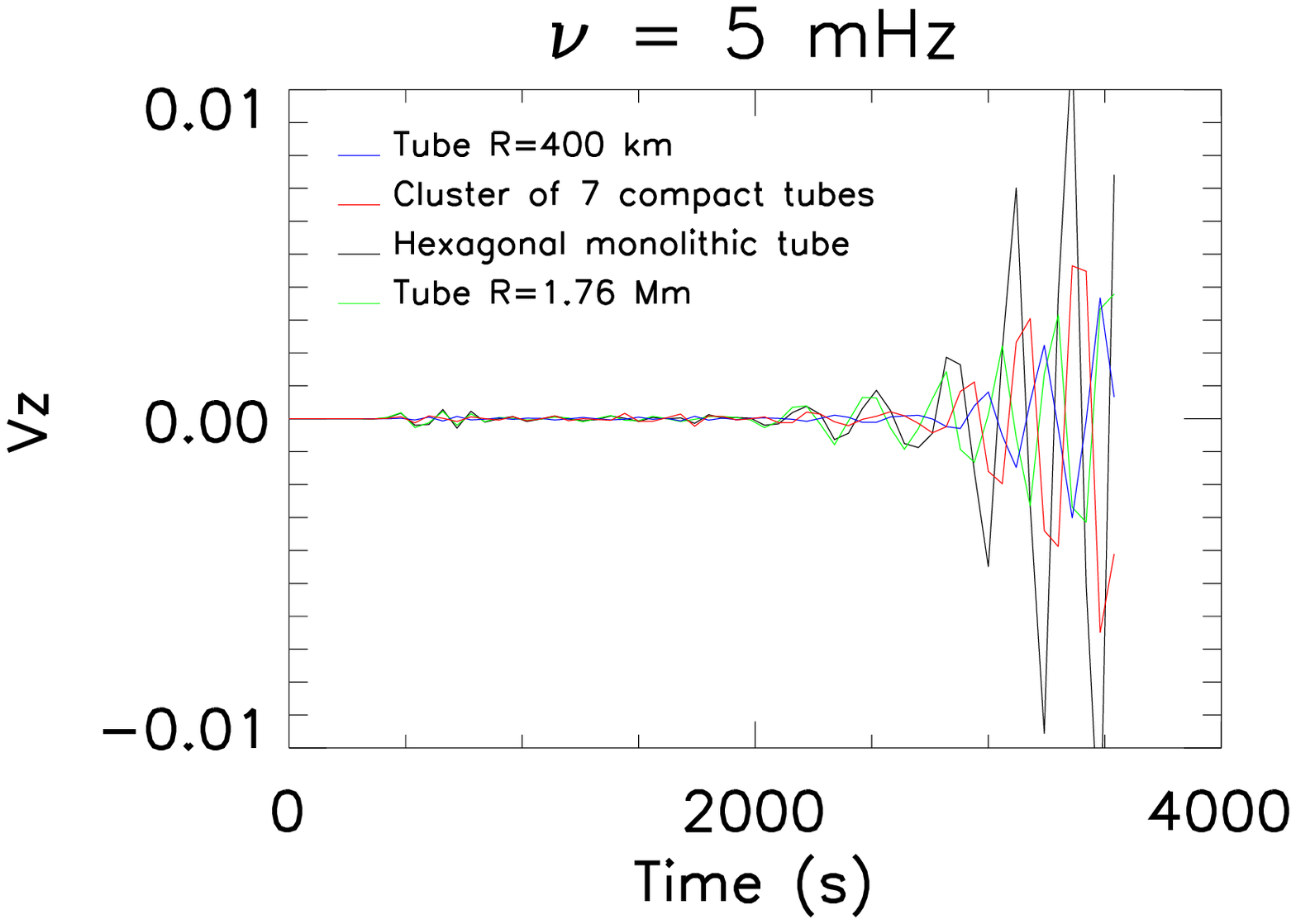}
 \caption {Scattered vertical velocity $V_z$ as a function of time measured at point B for the different magnetic features (A), (B), (C) and for a single monolithic tube of 400 km radius. The curves are plotted for the $f$-mode frequencies 2 mHz, 3 mHz, 4 mHz, and 5 mHz. } 
\label{fig6}
\end{figure*}


\section{$f$-mode Interaction with Large Size Sunspot Models}
     \label{S-largespots} 

In this section, we want to know if the multiple-scattering regime occurs for a larger size compact cluster, which corresponds to a larger separation distance ($d$). The distance $d$ in the case of a cluster is fixed as $d=3.4R_c$, where $R_c$ is the radius of an individual tube within the cluster. Nevertheless, multiple-scattering regime occurs if $d$ is about $ \lambda/2\pi \le d \le \lambda/2$. In this case a radius of $R_c = 400$ km satisfies the condition of multiple-scattering at the standard frequency $\nu=3$ mHz. Given that, we have made   
the simulations displayed in Figure \ref{fig5} where snapshots (A), (B) and (C) show the propagation of an $f$-mode wave packet through:\\

\begin{itemize}
\item  (A) a cluster of seven identical flux tubes in a hexagonal compact configuration. Each individual tube within the cluster has a radius $R_c=400$ km (Figure \ref{fig2} to the bottom),
 \item  (B)  a single monolithic tube whose radius $R$ = 1.76 Mm is the average radius of the cluster in (A), 
 \item  (C) a hexagonal monolithic flux tube which the shape at the surface is the same as that of the cluster in (A).
\end{itemize}

We observe from Figure \ref{fig5} that unlike the case of small sunspot models, the scattered near fields of the large hexagonal monolithic and monolithic tube models are different. The close near field of the hexagonal monolithic model (C) shows a non-uniform waveform, which indicates a contribution from the outer extents (or from the gaps between the extents) to the scattering.  

The cluster in snapshot (A) shows a triangular shape waveform in the left near field. This can be the result of oscillations in $x$-direction as well as in $y$-direction from tubes within the cluster under the multiple-scattering regime. 


\subsection{Multi-Frequency Effects on the Scattering}
     \label{S-flargespots} 
 
Figure \ref{fig6} shows the scattered vertical velocity $V_z$ as a function of time measured at point B (Figure \ref{fig5}) for the different magnetic features and for $f$-mode frequencies. As in the  Figure \ref{fig4}, the incoming wave in an unperturbed field  have an amplitude of 1. Unlike the small cluster model, we observe that the scattered curves of the large cluster are not in phase with that of the single tube of 400 km for all frequencies. 
Similarly, the scattering curves of the hexagonal monolithic model and its equivalent monolithic tube are no longer in phase, which indicates a different behaviour for both models. 

It is also observed that the scattering from the hexagonal monolithic model increases with frequency. It reaches a maximum at $\nu=4$ mHz and $\nu=5$ mHz where it dominates the scattering from the other models.


\section{Measurement of the scattering cross sections for the small and the large sunspot models}
\label{scatc}
In previous sections, we measured the scattered surface vertical velocity from a single point in the far field (point B). The obtained results can be checked with the observation of the solar surface with Dopplergrams. However, we need more quantitative analysis to qualify the scattering, not only from  a single point (B) but over all the $y$-ridge.  In this section, we use the vertical scattered wave, measured along the line $x=x_B$, to compute the scattering cross section $\sigma_{sc}$ for the different sunspot models and frequencies. 

The scattering cross section is a very important parameter in the scattering problem. It is defined for an incident plane wave as the total scattered power over the power per unit area of the incident wave. In our case, for a given frequency, the one-dimensional scattering cross section $\sigma_{sc}$ can be expressed as 

\begin{equation}
   \sigma_{sc}= \frac{\sum_y \sum_t |V_z^s(x_B,y,t)|^2}{\sum_t |V_z^0(x_B,y,t)|^2/d_B},
	\label{scatcross}
\end{equation}

where $V_z^s$ and $V_z^0$  are respectively the amplitudes of the scattered and the incoming vertical velocity at the surface measured at the point $x_B$. The distance $d_B$ is the separation between the center of the sunspot model and the point B along the $x$ axis ($d_B=7$ Mm).

\begin{figure*}  
\centering
\includegraphics[width=0.44\textwidth]{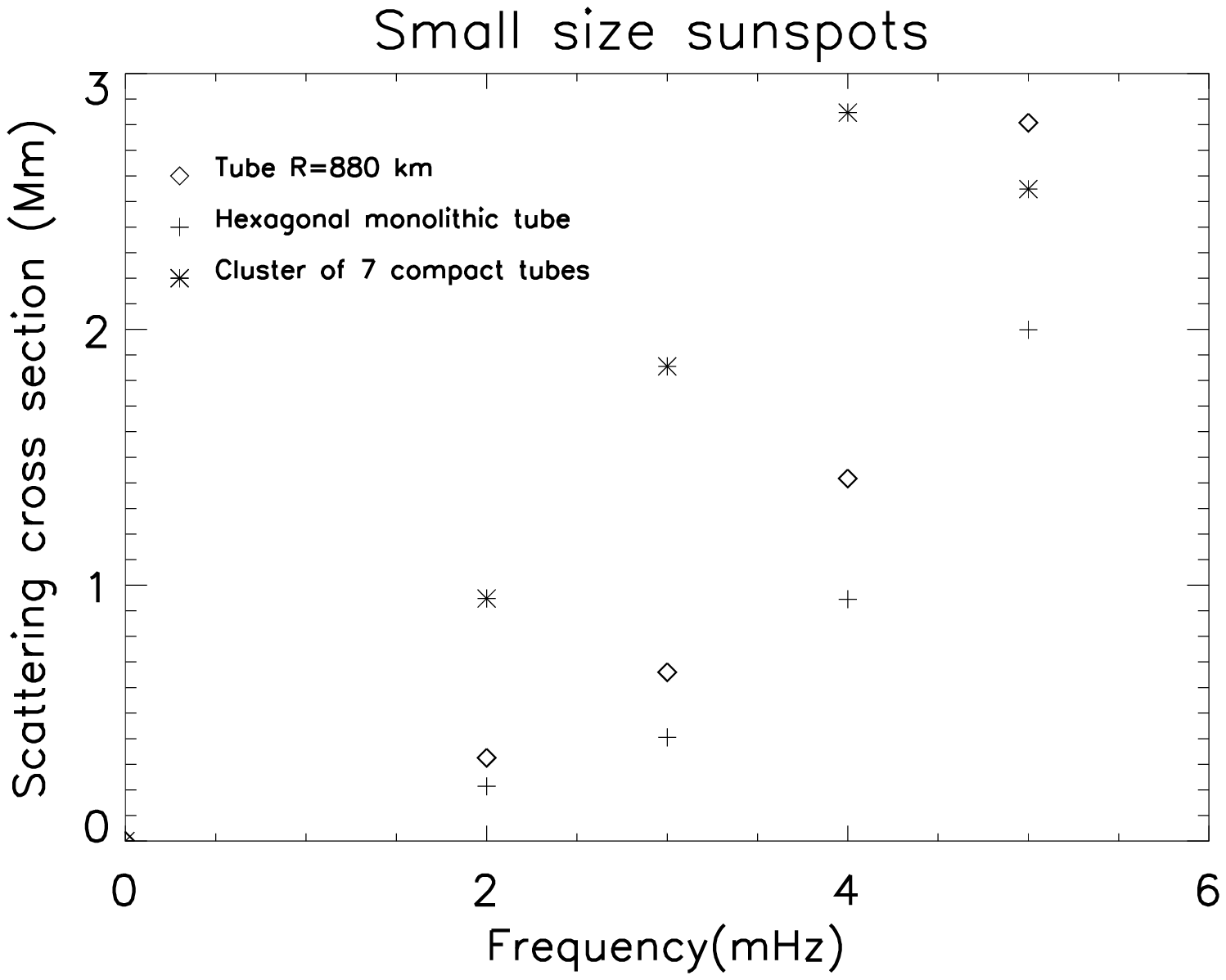} 
\includegraphics[width=0.44\textwidth]{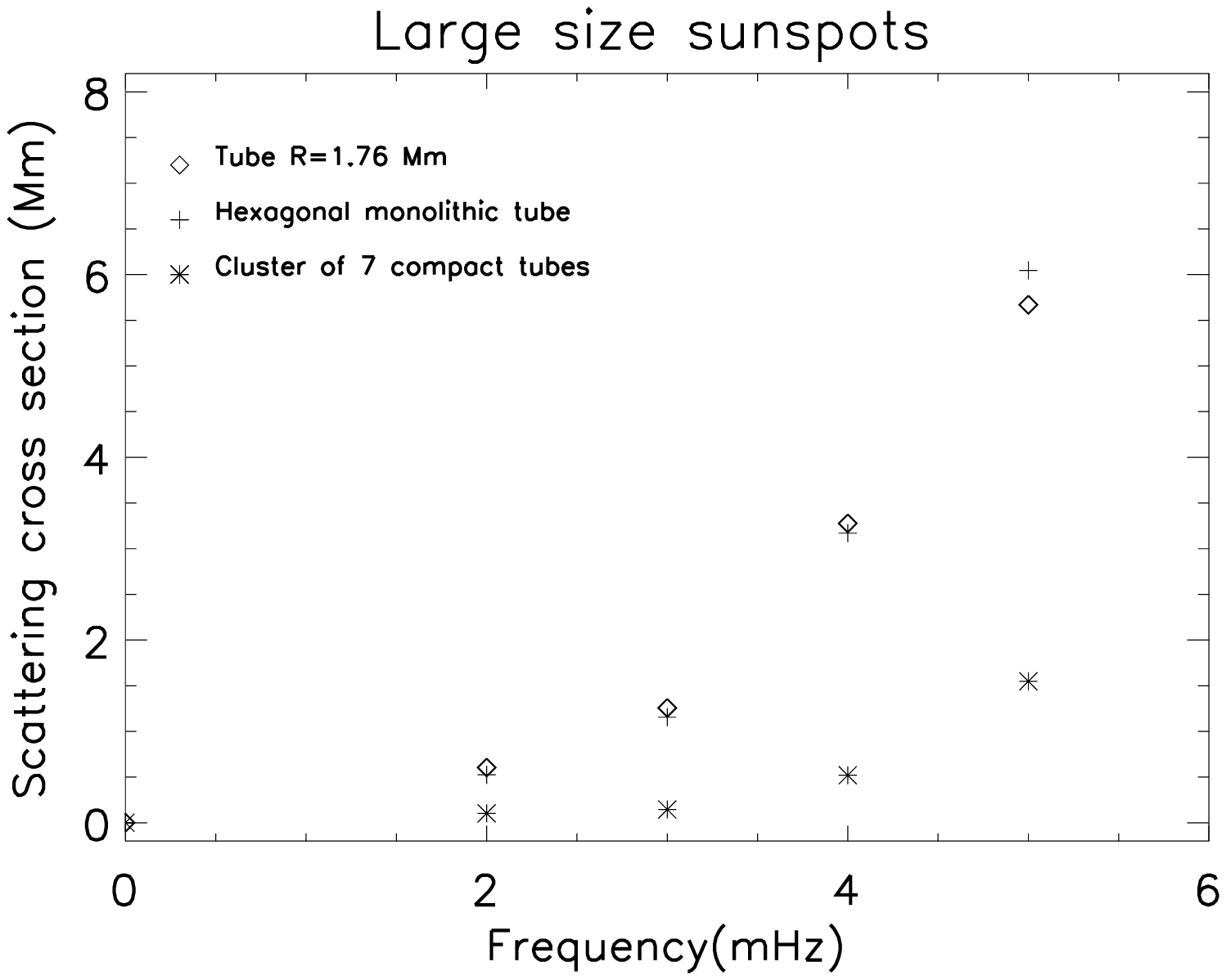}
 \caption {The scattering cross section for the small (left panel) and the large (right panel) sunspot models computed from the simulations.} 
\label{fig7}
\end{figure*}

Figure \ref{fig7} shows the scattering cross section computed for the small  and the large sunspot models. $\sigma_{sc}$ is computed over $y=0-192$ (from $y=-20$ Mm to $y=20$ Mm)  and from $t=0$ to $t=3300$ s. 

While small monolithic tube and hexagonal monolithic models oscillate in the same ways (Figure \ref{fig4}), the left panel of Figure \ref{fig7} shows that the scattering cross section for the small monolithic tube is larger than that of the small hexagonal model.

The scattering cross section of the small compact cluster model increases with the frequency over that of the two other models, except at the frequency of $\nu=5$ mHz where it decreases below that of the monolithic tube model. 
We attribute this particular behaviour to the absorption caused mainly by the simultaneous oscillations of tubes within the cluster in the $y$-direction.

Unlike the small cluster model, the large compact cluster model in the right panel of Figure \ref{fig7} shows a minimum of scattering cross section compared to the scattering from the other models. This result is explained by the multiple-scattering regime ($\nu=3, 4$ mHz) and the absorption from tubes within the cluster.

The large monolithic tube and hexagonal monolithic models have almost the same scattering cross section for the frequencies $\nu=2,3,4$ mHz, whereas at the frequency $\nu=5$ mHz, the scattering cross section of the large hexagonal monolithic model increases slightly above that of the monolithic tube model. This result can be seen clearly in Figure \ref{fig6} at the frequency $\nu=5$ mHz where the scattering from this model is larger than the scattering from the other models.

Actually, the geometrical shape of the large hexagonal model imposes a constraint on the oscillation modes that are excited in it in comparison with the cylindrical shape of the monolithic tube. In this case, more oscillation modes are excited inside the monolithic tube than inside the hexagonal model. Therefore, we have less absorption and more scattering of waves from the hexagonal model than from the monolithic tube model of the same size. This is a very important observation which reveals how the geometrical shape affects the input and output of waves in large sunspots particularly for a high frequency.

\section{Discussion and Conclusions}
\label{concl} 

In this paper, we are interested in finding a way to distinguish between distinct models that characterize the magnetic structure of sunspots. This can be observed through the scattered wave field when waves interact with these features.
Direct numerical simulations have begun to describe the scattering regime for an ensemble of magnetic flux tubes (e.g. \citealt{daiffallah14}). In the latter paper, we have studied the interaction of an $f$-mode of 3 mHz frequency with monolithic and cluster models of the sunspot. In this present study, we have incorporated for the first time a non-circular, or a hexagonal monolithic tube as a third model that can be used as a junction between the previous two models to better interpret the scattered wave field. 
While the truly sunspot like the latter model or the compact hexagonal cluster could not realistically exist, it is always useful as a first step to understand a minimum of physics or observations using simple models which do not require  important or expensive computational resources.

To discuss the scattering as a function of $d/\lambda$, instead of changing the distance between two neighboring tubes within the cluster $d$ as in \citet{daiffallah14}, we have fixed $d$ as in hexagonal compact cluster separation, and we have changed the wavelength $\lambda$ of the incoming wave through the variation of the frequency to see if we can have a multiple-scattering regime and absorption from a cluster in a close-packed configuration as in the case of a loose cluster in the same regime.

For more general results, we have performed simulations with two kinds of a cluster: 

1- Small cluster as in \citet{daiffallah14} which it is made of seven compact tubes of 200 km radius, where the separation distance $d=680$ km,

2- Large size cluster composed of seven compact tubes of 400 km radius, where $d=1.36$ Mm. 

For the frequencies $\nu=2,3$ mHz, we have demonstrated that the small cluster ($d \le \lambda/2\pi$) oscillates more like individual tube of 200 km than like monolithic tube of the same size, but with a larger amplitude. At this frequency range,  the scattering cross section of the small cluster is the largest compared to that of the other models, revealing that this model acts more like a scatterer object under these conditions.
This important result can be verified with helioseismic measurements to distinguish between close-packed and loose configurations of magnetic flux tubes inside sunspots or plage in part, and between fibril and monolithic configurations of sunspots in other part.

For the high frequency of 5 mHz, the small cluster which is supposed to be in a scattering regime ($\lambda/2\pi < d < \lambda/2$) oscillates in a different way compared to the other models. However, no signature of a multiple-scattering has been observed in the near field. Nevertheless, a distortion of the scattered wave field in the $y$-direction has been observed for this model. A similar observation was mentioned by \citet{felipe14} showing a more flattened scattered wave from a spaghetti model. We think that this particular signature is caused by the simultaneous oscillation in $y$-direction of the pairs of tubes aligned in a perpendicular direction to the incoming wave, independently from the scattering regime or the separation distance as shown by \citet{daiffallah14}. This effect combined with the multiple-scattering condition at this frequency can explain the absorption by the cluster observed at this specific frequency in both scattering amplitude and cross section plots.  This result constitutes an another criterion to distinguish a compact fibril sunspot from a monolithic one in this frequency range. We have to note that this effect should be amplified with the increasing of tubes number inside the cluster.

In contrast to the small cluster, the large size cluster shows a multiple-scattering in the near field and a minimum of scattering cross section, which indicates more absorption by tubes within the cluster. However, this absorption seems to be not significant to be observed in the scattering cross section plot as in the case of small sunspot model.

To have some scattering regime, tubes within cluster have to exchange their scattering through the separation distance $d$. In the case of a cluster of compact tubes, the distance $d$ is completely immersed in the magnetic field of the pair of tubes, where for a loose cluster, a part of the space between tubes is outside magnetic field. Therefore, somehow the magnetic field within the cluster of compact tubes does not support the scattering exchange between tubes in the horizontal direction, but rather supports simultaneous motion of tubes acting as a glue that holds the tubes together. Curiously, this characteristic describes the acoustic jacket phenomenon \citep{bogdan95} which is a near field of slow waves around the tube that propagate vertically in a stratified atmosphere carrying energy away, but they are evanescent in the radial direction. We know from previous studies that the gravitational stratification removes resonant absorption of a bundle of magnetic flux tubes and may reduce strong interactions between closely spaced flux tubes. \citet{hanasoge09} found that the mutual induction of the near-field jackets of two tubes on close separations can dramatically alter the scattering properties of the system playing an important role in the multiple-scattering regime. Given that, it is possible that the interaction between jacket modes of tubes within the cluster of compact tubes ($d < \lambda/2\pi$) inhibits the multiple-scattering regime. 

In conclusion, in addition to the minimum condition $d \sim \lambda/2\pi$ for the small cluster to have a  multiple-scattering regime, the distance $d$ between tubes within this model has to be larger than 
the distance between the locations of the minimum of the magnetic field strength in the pair of tubes.

The case of the large cluster model is different. It is possible that the scattering from larger size  individual tubes within this cluster is not completely absorbed through the jacket mode phenomenon, which explains probably the observation of near-field waves from this model despite  $d < \lambda/2\pi$.

In this context, we have to note that since this lower limit of multiple-scattering regime depends on the  size of the tubes, this distance would be much smaller in the case of the thin flux tube approximation.

Independently from the cluster model, our simulations show that the small size hexagonal monolithic model oscillates like its equivalent monolithic tube model for all frequencies, with more scattering for the latter model.

A more interesting behaviour is observed for the large size hexagonal monolithic model. This model and its equivalent monolithic tube have approximately the same scattering cross section for low and mean frequencies. However, the large hexagonal model shows less absorption in a high frequency. In fact, due to its geometrical shape, less oscillation modes are excited inside compared to the supported waves in the monolithic tube model. This is an important result, demonstrating a reasonable effect of the sunspot geometrical shape on the interaction of high-frequency waves with large monolithic models, which must be taken into consideration in addition to the radius and the wavelength in future simulations.

This work is a step toward the understanding of the helioseismic signature of sunspot models. More improved numerical and analytical investigations are necessary to better interpret the observations.


\section*{Acknowledgements}
\addcontentsline{toc}{section}{Acknowledgements}

The author thanks Robert Cameron and Toufik Abdelatif for useful discussions.
The author also thanks the anonymous referee for constructive comments and suggestions that improve the quality of the paper.








\bsp	
\label{lastpage}
\end{document}